\documentclass[11pt,reqno]{amsart}
\usepackage{amsmath}
\usepackage{amsthm}
\usepackage{float}

\usepackage{upgreek}
\usepackage{verbatim}
\usepackage{hyperref}
\usepackage{mathrsfs}
\usepackage{bm}
\usepackage{dsfont}
\usepackage{amsfonts}
\usepackage{amssymb}
\usepackage{csquotes}
\usepackage{stmaryrd}
\usepackage{mathtools,bbm}
\usepackage[font=small,labelfont=bf]{caption}
\usepackage{subcaption}

\usepackage{enumerate}

\usepackage{soul}
\usepackage[margin=1in]{geometry}
\usepackage{graphicx}
\usepackage{xcolor}

\numberwithin{equation}{section}

\begin{document}
\title{A Macroeconomic SIR Model for COVID-19}
\author[E. Bayraktar]{Erhan Bayraktar \address{(E. Bayraktar) Department of Mathematics, University of Michigan, Ann Arbor, Michigan 48109, United States}\email{erhan@umich.edu}}
 
 \author[A. Cohen]{Asaf Cohen \address{(A. Cohen) Department of Mathematics, University of Michigan, Ann Arbor, Michigan 48109, United States}\email{asafc@umich.edu}}

 \author[A. Nellis]{April Nellis \address{(A. Nellis) Department of Mathematics, University of Michigan, Ann Arbor, Michigan 48109, United States}\email{nellisa@umich.edu}}

\date{\today}

\maketitle
\begin{abstract}
The current COVID-19 pandemic and subsequent lockdowns have highlighted the close and delicate relationship between a country's public health and economic health. Macroeconomic models which use preexisting epidemic models to calculate the impacts of a disease outbreak are therefore extremely useful for policymakers seeking to evaluate the best course of action in such a crisis. We develop an SIR model of the COVID-19 pandemic which explicitly considers herd immunity, behavior-dependent transmission rates, remote workers, and indirect externalities of lockdown. This model is presented as an exit time control problem where the lockdown ends when the population achieves herd immunity, either naturally or via a vaccine. A social planner prescribes separate levels of lockdown for two separate sections of the adult population - those who are low-risk (ages 20-64) and those who are high-risk (ages 65 and over). These levels are determined via optimization of an objective function which assigns a macroeconomic cost to the level of lockdown and the number of deaths. We find that, by ending lockdowns once herd immunity is reached, high-risk individuals are able to leave lockdown significantly before the arrival of a vaccine without causing large increases in mortality. Additionally, if we incorporate a behavior-dependent transmission rate which represents increased personal caution in response to increased infection levels, both output loss and total mortality are lowered. Lockdown efficacy is further increased when there is less interaction between low- and high-risk individuals, and increased remote work decreases output losses. Overall, our model predicts that a lockdown which ends at the arrival of herd immunity, combined with individual actions to slow virus transmission, can reduce total mortality to one-third of the no-lockdown level, while allowing high-risk individuals to leave lockdown well before vaccine arrival.
\end{abstract}

\section{Introduction}
The current COVID-19 global pandemic has led to massive lockdowns to slow the spread of the virus. Now, policymakers face a dilemma - extended periods of lockdown have put strain on the economy, but returning to ``normal'' too quickly could result in an equally troubling second wave of infections. The task is therefore to find the optimal balance between public health and economic growth. Models such as those proposed by Alvarez et al. \cite{alv} and Acemoglu et al. \cite{acem} have used a macroeconomic approach and variations on the Susceptible-Infectious-Recovered (SIR) epidemic model proposed by Kermack et al. \cite{sir} to solve an optimization problem determining the lockdown policy that minimizes both loss of life and effects on output. We consider a variation on these models which incorporates several new concepts and gives a wider picture of the overall situation. 
\begin{itemize}
    \item We formulate an exit time control problem where lockdown measures are lifted when the population reaches herd immunity, even if this occurs before a vaccine is developed.     
    
    \item We incorporate a transmission rate that captures how individuals reacts to current infection levels, as discussed in \cite{cochrane}. This ``behavior-dependent'' transmission rate seeks to model individual behaviors that occur independently of lockdown. For example, individuals might wear masks, practice social distancing, and take other precautions to reduce their risk as infection numbers go up, even in the absence of official lockdown measures.     
    
    \item We consider the costs of indirect deaths attributed to adverse mental and physical effects of lockdown and of continued unemployment after the lockdown has ended, and the positive impact of workers who are able to work remotely during lockdown.  
       
    \item We add a penalty for overwhelming intensive care unit (ICU) capacity and a term that captures the future impacts of missed health screenings during the pandemic.     
\end{itemize}

The planning problem developed in \cite{alv} is the basis of the one used in our model. The paper references the SIR method of epidemic modeling, which is also used in subsequent papers, to represent population dynamics. Many of its parameters, like level of obedience of lockdown, are also used in our model. The death rate is calculated as a function of the number of infected individuals in order to model the effects of hospital overcrowding and encourage ``flattening the curve''. The objective function quantifies the economic and social impacts of both the pandemic and the resulting lockdown measures, and develops an optimization problem for a planner to solve. The cost of lockdown is represented by the income that is lost by those who are in quarantine and so are unable to work, while the cost of death is calculated as value of statistical life. Their model examines the role of the population's level of obedience, as well as the effect of being able to test those who are recovered and exclude them from lockdown. They also investigate the results of different values of statistical life. The authors conclude that being able to test and return recovered individuals to the workforce has a large positive effect on outcomes and that these outcomes are sensitive to the fatality rate and its elasticity with respect to infection level. 

A subsequent paper, \cite{acem}, takes this model and extends it by considering the possibility of different optimal lockdown measure for different groups. In their case (and in ours) the groups are differentiated by age, since the severity of COVID-19 infection varies widely based on age. The paper also explicitly considers the number of infected individuals admitted to the ICU at each point in time, which is then used to calculate the death rate. The authors use Pareto curves created by varying the non-pecuniary value of life to show that targeted lockdown measures unilaterally perform better than uniform lockdowns, regardless of whether one seeks to prioritize reducing output loss or reducing mortality. In fact, while they consider three age groups (20--49, 50--64, and 65+), their results show that it is sufficient to consider a ``semi-targeted'' policy which prescribes one lockdown policy to those aged 20--64 and another policy for those over 65 years of age. Due to this result, we also split the working population into two groups, one aged 20--64 and one aged 65 and over.

We are also influenced by the work of Cochrane \cite{cochrane}, which discusses a Behavioral SIR (BSIR) model wherein individual behaviors affect infection transmission rates. This reflects the tendency of individuals to be more careful as infection levels rise in their community. In this model, the dynamics eventually reach a stable equilibrium with a virus reproduction rate equal to 1 and a nonzero constant rate of infections and deaths per day. To handle this, he considers the role of technologies like testing and tracing, which could allow people to reduce transmission levels while still maintaining economic activity. We adopt the idea of behavior-dependent transmission based on infection levels, but instead rely on lockdown and the arrival of herd immunity or a vaccine to drive infection numbers to zero. 

These works develop a solid framework for our model, but we undertake the task of increasing accuracy and realism through herd immunity as an exit time, behavior-dependent transmission rates, deaths indirectly due to lockdown, additional costs of lockdown, and the portion of the population that is able to work remotely. These additions affect the model in various ways, but  overall our augmented model concludes that the high-risk group can be released from lockdown before a vaccine arrives \textit{without} large adverse effects (though of course, this is not to say that a vaccine is not necessary or useful). 

\begin{itemize}
\item When the expected vaccine arrival time is 1.5 years after the start of the outbreak, our model recommends less than 7 months of lockdown for the high-risk group (instead of locking down for the full 1.5 years until the vaccine). Additionally, lockdowns for the low-risk group are 6 weeks shorter.
 
\item The addition of a behavior-dependent virus transmission rate contributes to these shorter lockdowns and decreases mortality. In an extreme situation where individuals can take measures that decrease transmission by 95\% when infections reach 30\%,  less than a month of lockdown is prescribed for the low-risk group. In the more moderate benchmark case, where individuals are able to reduce their transmission by 25\% when infections reach 30\% of the population, herd immunity arrives a month earlier than in a situation with a constant disease transmission rate. In both cases, we also observe lower output loss due to shorter lockdown and fewer deaths due to slower transmission.

 \item Increasing the level of remote work reduces the impact of COVID-19 by decreasing both mortality and output loss, even though a longer lockdown is imposed. This supports the intuitive idea that increased remote work reduces infection risk without sacrificing economic activity. 
 
 \item Increasing the predicted length of future unemployment and the predicted rate of lockdown-related deaths both decrease lockdown length in a similar manner, but also have negative impacts on outcomes. Adjusting the length of future unemployment and the predicted number of indirect deaths due to lockdown lead to trade-offs between output and mortality.  Running the model with different initial conditions shows that higher pre-lockdown infection levels lead to earlier onset of herd immunity but higher death tolls, highlighting the risks of infection spikes. Future impacts of current missed health screenings and a penalty for overfull ICUs are revealed to have little impact on the optimal lockdown policy, at least in our formulation. 
\end{itemize}

The main body of the paper presents our model and its numerical results. In Section \ref{sec:pop-dyn}, we lay out the SIR dynamics used to model the transmission of the virus and discuss certain model additions, especially the addition of deaths indirectly caused by lockdown and a behavior-depending transmission rate. In Section \ref{sec:obj-func}, we introduce the exit time control problem that ends when the population reaches herd immunity and discuss the terms in the objective function. In Section \ref{sec:numerics}, we discuss our numerical model, which discretizes the problem and is solved through value iterations. We calibrate it with the results of \cite{acem} and \cite{alv} and compare these results to our augmented model using death rates on the same scale. Then, we update death rates to match more recent data from \cite{CDCstats} and adjust the non-pecuniary value of life. We present and discuss our results and perform some parameter robustness analysis. These experiments serve to illustrate the general mechanisms of the model and to present planners with an idea of our model's potential. If a planner wishes to use our model, parameter values can be changed in our code, found at \footnote{\url{https://github.com/april-nellis/COVID19-BSIR}}, to accurately reflect a specific planner's current situation.

\section{Population Dynamics}
\label{sec:pop-dyn}

As in \cite{acem}, we consider policies which assign different lockdown strategies to population groups with different responses to infection and lockdown. Influenced by their results, we divide adults into one group aged 20-64, called``low-risk" and indexed by $j = 1$, and one group aged 65 and over, called ``high-risk" and indexed by $j=2$. We will only consider adults older than 20, so the low-risk group makes up 82\% of the population of interest, while the high-risk group makes up 18\% \cite{census}. The second group can also include individuals of any age who are more likely to contract severe cases of COVID-19 and experience complications due to immunodeficiency, respiratory weakness, or other preexisting conditions. These individuals are considered separately from the general working population. We denote the population of group $j$ as a proportion, $N_j$, of the total population and lay out the following relationship:
\begin{equation*}
    S_j(t) + I_j(t) + R_j(t) + D_j(t) = N_j,\ j \in \{1, 2\},\ \text{ where } \sum_j N_j = 1.
\end{equation*}

Individuals move from susceptible ($S_j(t)$) to infected ($I_j(t)$) to recovered ($R_j(t)$), with additional flows from all three to death ($D_j(t)$). The dynamics are given by

 \begin{align}
 \begin{split}
      &\dot S_j(t) = - \dot I_j(t) - \gamma I_j(t) - \xi(L_j(t))S_j(t), \\ 
     &\dot I_j(t) = S_j(t)(1 - \theta L_j(t))\sum_k \beta_{kj}(t)(1 - \theta L_k(t))I_k(t) - \gamma I_j(t), \\
     &\dot R_j(t) = \gamma I_j(t) - \phi(I_j(t))I_j(t) - \xi( L_j(t))R_j(t), \\
     &\dot D_j(t) = \phi(I_j(t)) I_j(t) +\xi(L_j(t))(S_j(t) + R_j(t))= -\dot N_j(t).
     \label{eq:dynamics}
\end{split}
 \end{align}
 
The number of new infections depends on the size of the susceptible and infected populations, as well as the lockdown levels, $L_j(t)$, and transmission rates between groups, $\beta_{ij}$. As others have suggested (\cite{acem}, \cite{alv}), a portion of the population will disregard lockdown orders. This level of obedience is represented by $\theta$. As in \cite{acem}, we set $\theta= 0.75$, though this parameter is difficult to quantify exactly. Patients move out of the infected category with rate $\gamma$ in accordance with the expected recovery time of 18 days. Deaths due to COVID-19 occur at rate $\phi(I_j(t))$ and other deaths occur at rate $\xi(L_j(t))$. For convenience, the parameters that appear in \eqref{eq:dynamics} and in the objective function \eqref{eq:obj}, along with their levels, are listed in Table \ref{tab:param} of the appendix. 
 
 \subsection{Deaths}
 One unique element of COVID-19 is its significantly different death rates for different groups. Therefore, the base death rate $\delta^0_j$ for each group is set individually. In addition, as the number of infected individuals increases and hospitals become more crowded, death rates increase as a function of the total number of infected patients as in \cite{alv}. We represent this increase in the death rate as $\delta^1_j$ and follow \cite{acem} in assuming that an infection level of 30\% increases the death rate by a factor of five. Therefore, the death rate due to viral infection is
  \begin{equation*}
     \phi(I_j(t)) = \delta^0_j + \delta^1_j \sum_j I_j(t).
 \end{equation*}
 
Additionally, as lockdowns stretch on concerns have been raised regarding ``deaths of despair'' due to the impacts of lockdowns on mental health (\cite{despair1}, \cite{despair2}). Hospitals have also shut down many departments to accommodate the increased need for ICU units for COVID-19 patients. Many non-elective surgeries and routine health checks have also been cancelled or rescheduled (\cite{screenings}). This neglect of health maintenance is also very likely to have repercussions on public health. To encompass all this, we add the term $\xi(L_j(t))$, which is written as
 \begin{equation*}
     \xi(L_j(t)) = \alpha_L L_j(t).
 \end{equation*}
 This represents the number of deaths indirectly caused by the lockdown, and scales with $L_j(t)$. We argue that, in the absence of any way to verify immunity, indirect deaths occur in both the susceptible and recovered populations. So, the total number of deaths is given by
\begin{equation*}
    \sum_j \phi(I_j(t))I_j(t) + \xi(L_j(t))(S_j(t) + R_j(t)).
\end{equation*}

\subsection{Behavior-dependent disease transmission}
The basic transmission rate of COVID-19 is approximately 0.2 (\cite{alv}, \cite{acem}). This means that about 20\% of those who come in contact with an infected individual will become infected themselves. However, we incorporate a transmission rate that decreases as infections increase due to increased caution between people, as discussed in \cite{cochrane}. In addition, we consider an inter-group interaction factor, $\rho$, as in \cite{acem}. This reflects a lower rate of interactions between groups. For example, working people aged 20-64 will interact more with their peers than with those in the high-risk group. Combining these two ideas, we represent the transmission rate as
\begin{equation*}
    \beta_{kj}(t) = \begin{cases} \rho\beta_0e^{-\alpha_I I(t)} & \text{if } k \neq j, \\ \beta_0e^{-\alpha_I I(t)} & \text{if } k = j. \end{cases}
\end{equation*}
The scale factor $\alpha_I$ is chosen such that the rate of transmission can be decreased by a factor of $e^{-0.3\alpha_I}$ when 30\% of the population is infected with the virus. 

\section{Objective Function}
 \label{sec:obj-func}
 
 The questions of how long and how severely to lock down the population during a pandemic can be thought of as a planning problem. Given the above dynamics, we model the optimization problem that must be solved by a social planner using the following objective function, which represents the overall societal costs of a given lockdown policy: 
\begin{align}
\begin{split}
    \min_{L\in \Lambda} \int_{0}^{\sigma} e^{-(r+\nu) t} & \Big(\sum_j\Big[ \omega_j L_j(t)\big(S_j(t) + I_j(t) + pR_j(t)\big)(1 - h)  \label{eq:obj} \\
    & +(\chi + \frac{\omega_j}{r}(1 - e^{-r\Delta_j}))\phi_j (I(t))I_j(t)  + \frac{\omega_j}{r}(1 - e^{-r\Delta_j})\xi( L_j)(F + S_j(t) + pR_j(t)) \\
      &+  \alpha_E  \omega_j L_j(t)\big(S_j(t) + I_j(t) + pR_j(t)\big) \Big]  + \Omega(t) \Big)dt.
\end{split}
\end{align}

\subsection{Attainable Lockdown Levels} 
There are certain jobs that must be done even during a pandemic, preventing the population from attaining full lockdown. These essential professions include healthcare workers, grocery store employees, delivery workers, and the postal service, among others. Because of this, we set an upper limit on the possible lockdown level, denoted $\bar L_j$, and the set of possible lockdown policies is written as $\Lambda = [0,\bar L_1] \times [0, \bar L_2]$. We set $\bar L_1$at 0.7 to account for essential workers in the low-risk group. On the other hand, $\bar L_2$ is set at 1 since we assume that the high-risk group does not work.

\subsection{Herd Immunity}
Previous models considered either an infinite time horizon with stochastic vaccine arrival or a fixed horizon with deterministic vaccine arrival when formulating their objective function. We contribute a new approach that sets reaching herd immunity as the end of the problem. This can be reached either naturally via infection spread and recovery (which confers immunity) or via the arrival of a vaccine. We assume that those who have been infected with COVID-19 once will remain immune for the rest of the outbreak. Additionally, we assume that if a vaccine is approved for distribution, vaccination levels will be high enough to produce herd immunity. We define $\sigma$ as the time at which herd immunity is reached. We assume a herd immunity threshold of 60\% recovered\footnote{In \cite{CDCstats} Table 1 (Scenario 5: Current Best Estimates), the basic reproduction number of COVID-19 is $R_0 = 2.5$. Herd immunity is calculated as $1 - 1/R_0 = 0.6$.}, so we set $\sigma = \sigma(60) = \inf\{t\geq 0 : \sum_j R_j(t) \geq 0.6\}$.

\subsection{Output Loss}
The most noticeable result of lockdown measures is economic slowdown. Many workers who are not deemed essential and cannot work remotely have found themselves jobless as companies lose revenue. As in \cite{acem}, we take the average wage of a full-time worker and normalize it to 1 and assume that on average, those in the high risk group do not earn any wages. We do not assume the existence of an ``immunity passport'' given to those who have recovered and are immune, so we set a flag parameter $p = 1$ (this can be set to 0 to be consistent with the cases of \cite{alv} and \cite{acem}). On the other hand, we do consider some proportion $h$ of the workforce who are able to work from home. Therefore, we denote the purely salary-based cost of lockdown as
\begin{equation*}
    \omega_j L_j(t)(S_j(t) + I_j(t) + R_j(t))(1 - h).
\end{equation*}
When presenting our numerical results, we refer to the \textit{output} loss due to lockdown. This is not the value of the objective function presented in \eqref{eq:obj}, but rather the losses in output caused by requiring people to stay home and not work. The output loss is represented by
\begin{equation*}
    \int_{0}^{\sigma} e^{-(r+\nu) t} \sum_j \omega_j L_j(t)\big(S_j(t) + I_j(t) + pR_j(t)\big)(1 - h) dt
    \label{eq:output}
\end{equation*}
and is compared to annual ``normal'' output. This baseline output is calculated as the amount of output produced until the expected vaccine arrival time, $1/nu$, if there is no lockdown, annualized using the expected vaccine arrival time. This is given by 
\begin{equation*}
\nu \int_0^\infty e^{-(r + \nu)t} \sum_j w_j N_j dt = \dfrac{\nu}{r + \nu} \sum_j w_j N_j.
\end{equation*}

\subsection{Cost of Death}
We calculate the cost of a COVID-19 death in group $j$ in the same manner as in \cite{acem}. Here, $\chi$ is the non-pecuniary cost of life, which we consider as a measure of the \textit{public} impact of deaths due to COVID-19. This can be thought of as a measure of the planner's priorities. Lower values of $\chi$ lead to prioritizing output loss minimization, while higher values are chosen to encourage longer lockdowns and decrease mortality at the expense of output. To ensure that this cost is on the same order of magnitude as wages, we scale by the interest rate when we choosing $\chi$, similarly to \cite{alv}. Note that $\chi = 0.2/r$ is consistent with \cite{acem} where $\chi = 20$ and $r = 0.01$. $\Delta_j$ is the number of years left in an average individual's career. We set $\Delta_1 = 20$ and $\Delta_2 = 0$. Therefore, the cost per death due to COVID-19 is given by
\begin{equation*}
    \chi + \frac{\omega_j}{r}(1 - e^{-\Delta_j r}).
\end{equation*}

Deaths indirectly caused by the lockdown are not explicitly categorized and counted, and so can be considered ``invisible deaths''. For this reason, we do not include $\chi$ in the cost of these deaths and only count lost productivity. We also account for similar future deaths due to lack of preventative healthcare using a constant $F$ (the number of indirect deaths in the future relative to those that occur during lockdown). These deaths do not appear in the dynamics, as they have not yet occurred, but they are considered when calculating the costs of lockdown. For this reason, $F$ appears in \eqref{eq:obj} but not in \eqref{eq:dynamics}, and the total cost of indirect deaths is given by
 \begin{equation*}
\label{eq:indir}
     \frac{\omega_j}{r}(1 - e^{-r\Delta_j})\xi( L_j)(F + R_j(t) + S_j(t)).
 \end{equation*}
 
\subsection{Future Loss of Employment} Another addition to the model acknowledges the long-lasting economic impacts of a period of economic slowdown. Already some large corporations like JCPenney and Hertz have filed for bankruptcy (\cite{bankruptcy}), and there are certainly more companies, both large corporations and small business, who are under financial strain. Federal stimulus measures may alleviate some of this burden, but they cannot completely compensate for current drops in consumption. Effects may manifest in a variety of ways, but we choose to express them as a ``future loss in employment'', in which 1 day in lockdown results in some $\alpha_E$ days of lost employment (on average) after lockdown ends. We set this to be 0.42 (reflects current 14.7\% unemployment \cite{unemployApril20} and average 3 days of unemployment for one day of lockdown based on median unemployment duration of 25.2 weeks (6 months) in 2010 \cite{2008recession}). The cost of future unemployment is modeled by
 \begin{equation*}
    \alpha_E  \omega_j L_j(t)\big(S_j(t) + I_j(t) + pR_j(t)\big).
 \end{equation*}

 \subsection{ICU Overcapacity} 
 A major incentive for lockdown measures is ``flattening the curve'' -- slowing the spread of the virus so that hospitals and ICUs will not get overwhelmed by a flood of patients in need of ventilators and other specialized medical equipment. This is already reflected in the death rate which increases as infections increase, but we add an additional penalty on top of that. We assume that a fixed proportion of infected patients, $\iota_j$, require ICU care. We set this level to be 2.6\% for people without underlying conditions and 7.4\% for high-risk groups \cite{CDCstats}. Then, we incorporate a penalty $\eta$ (representing a daily penalty scaled by the level of overcapacity) for hospitalizations exceeding the estimated average ICU capacity, which is 30 beds per 100,000 people \cite{icu-capacity}. This is done via the function 
 \begin{equation*}
    \Omega(t) \coloneqq \eta\Big[\sum_j \iota_j I_j(t) - ICU\Big]^+  \times \max_j{\omega_j}.
 \end{equation*}
 
\section{Numerical Results}
\label{sec:numerics}

\subsection{Numerical Method}
We used value iterations, first introduced in \cite{bellman-57}, to solve the optimization problem presented by our model. The model is discretized using first-order Taylor approximations and the value function is calculated over a regular grid. Because the change in population due to deaths is very small, we follow the precedent set in \cite{alv} and iterate over a four-dimensional $(S_1, S_2, I_1, I_2)$ grid to determine optimal lockdown policy instead of the larger and more computationally expensive (but more accurate) six-dimensional grid $(S_1, S_2, I_1, I_2, R_1, R_2)$. By this, we mean that instead of separately keeping track of the recovered and dead populations, they are considered together as one unit when determining the optimal lockdown policy. Since the vast majority of this ``non-susceptible'' group is recovered, this simplification, which removes two state variables, has a small effect on accuracy but a large effect on computational complexity. And, when determining the pandemic trajectory for given initial conditions and a given lockdown policy, total deaths can still be calculated via the population dynamics shown in \eqref{eq:dynamics}. We choose $\Delta S = 0.0714$ and $\Delta I = 0.0357$ in our discretization, and set days as the unit of time. For all models, we take the initial conditions to be uniform across groups (if applicable) and set them at the level of 98\% susceptible, 1\% infected, 1\% recovered unless otherwise specified. 

\subsection{Calibration}
\label{sec:calib}
\begin{figure}
    \centering
    \includegraphics[width = \textwidth]{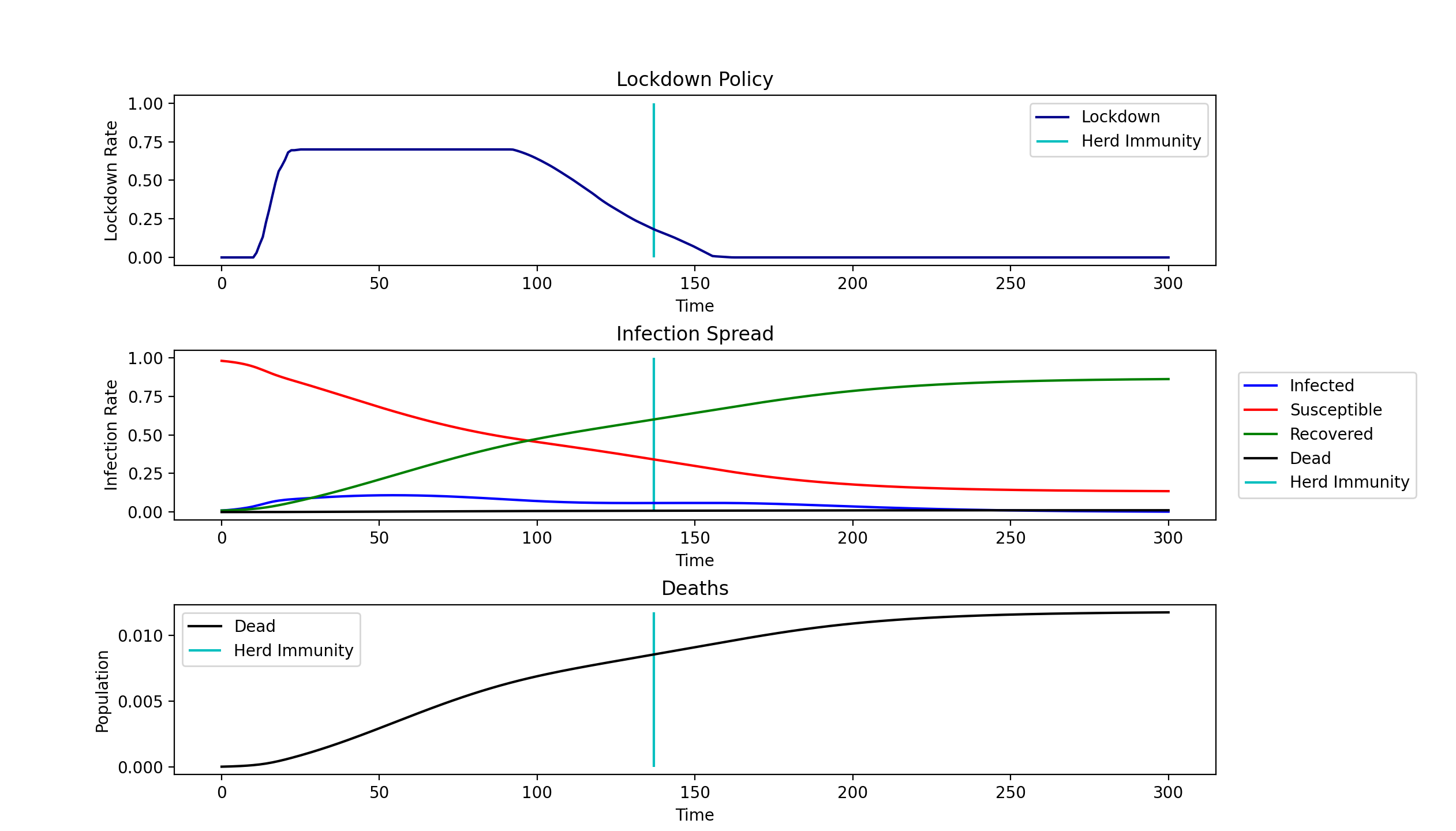}
    \caption{Recreation of \cite{alv} model (no groups or herd immunity), parameters from Table \ref{tab:param} \\ Output Loss: 13.4232\%, Total Deaths: 1.1754\%}
    \label{fig:alv}
\end{figure}

\begin{figure}
    \centering
    \includegraphics[width = \textwidth]{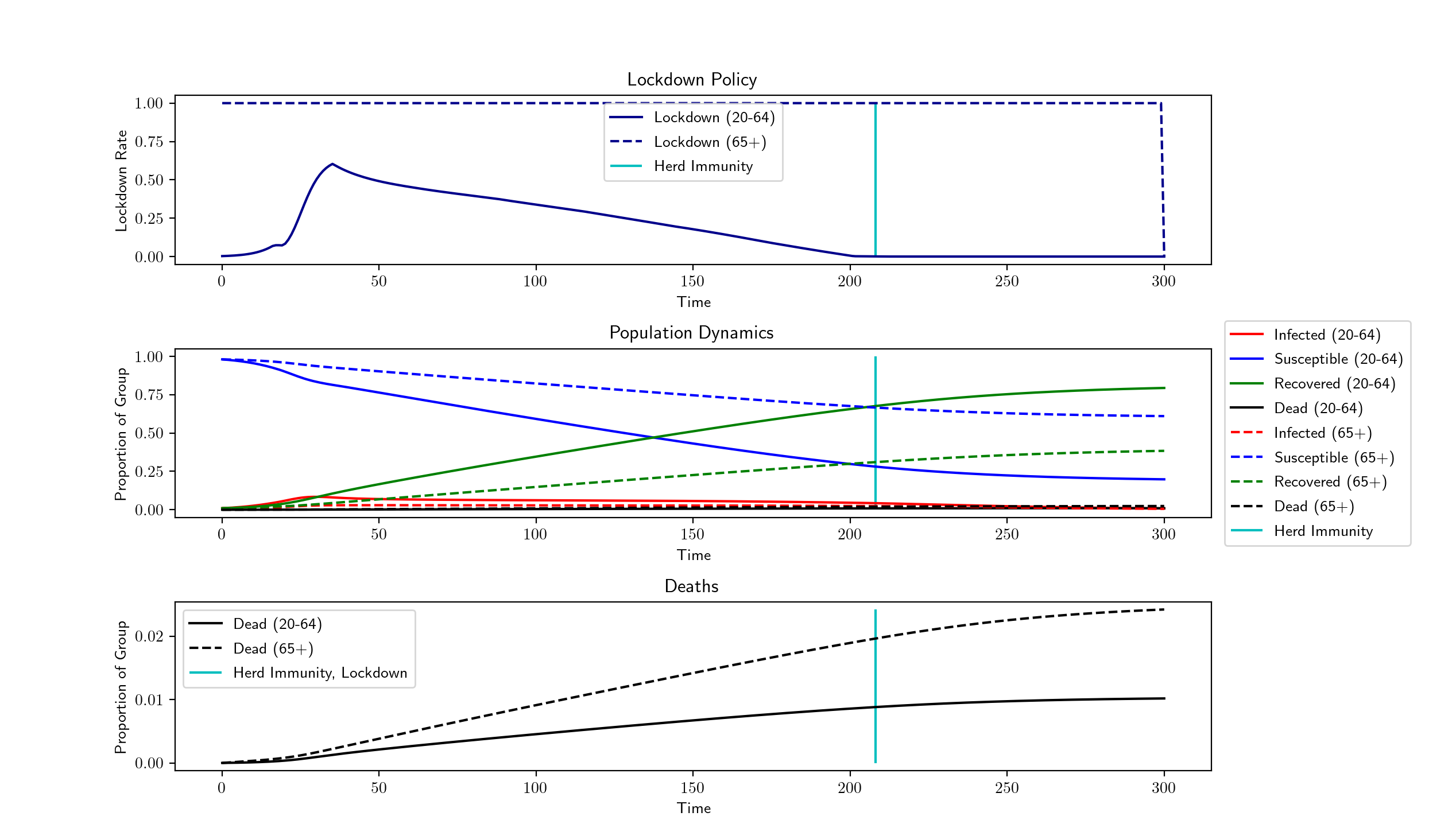}
    \caption{Recreation of \cite{acem} model (two groups and no herd immunity), parameters from Table \ref{tab:param} \\ Output Loss: 8.9676\%, Total Deaths: 1.3121\%}
    \label{fig:acem}
\end{figure}

To test the validity of our numerical models, we use the parameter values of \cite{alv} and \cite{acem} and compare our model's recommendations to their results. A full list of the parameter values used in this section is presented in Table \ref{tab:param}. In Figure \ref{fig:alv}, we compare a one-group version of our model with the one presented in \cite{alv}, whose recommended optimal lockdown reaches 70\% lockdown after about one month, and then slows reduces in intensity until lockdown is lifted approximately 140 days (4.5 months) after the outbreak begins. Our version of this model maintains the maximum lockdown of 70\% for slightly longer and ends slightly later. Interestingly, the ending of lockdown nearly coincides with the population reaching herd immunity, though we did not add any such considerations when running this example. In Figure \ref{fig:acem}, we set up our model to mimic the semi-targeted policy from \cite{acem} and find similar levels of output loss, total deaths, and general lockdown recommendations. Namely, the optimal strategy keeps the high-risk group in lockdown until the arrival of a vaccine, while the low-risk group is able to emerge and return to work after approximately 200 days of lockdown have elapsed. In this figure, note that the population reaches herd immunity well before the arrival of a vaccine, implying that the lockdown on the high-risk group could have been ended earlier.

\begin{figure}
    \centering
    \includegraphics[width = 0.95\textwidth]{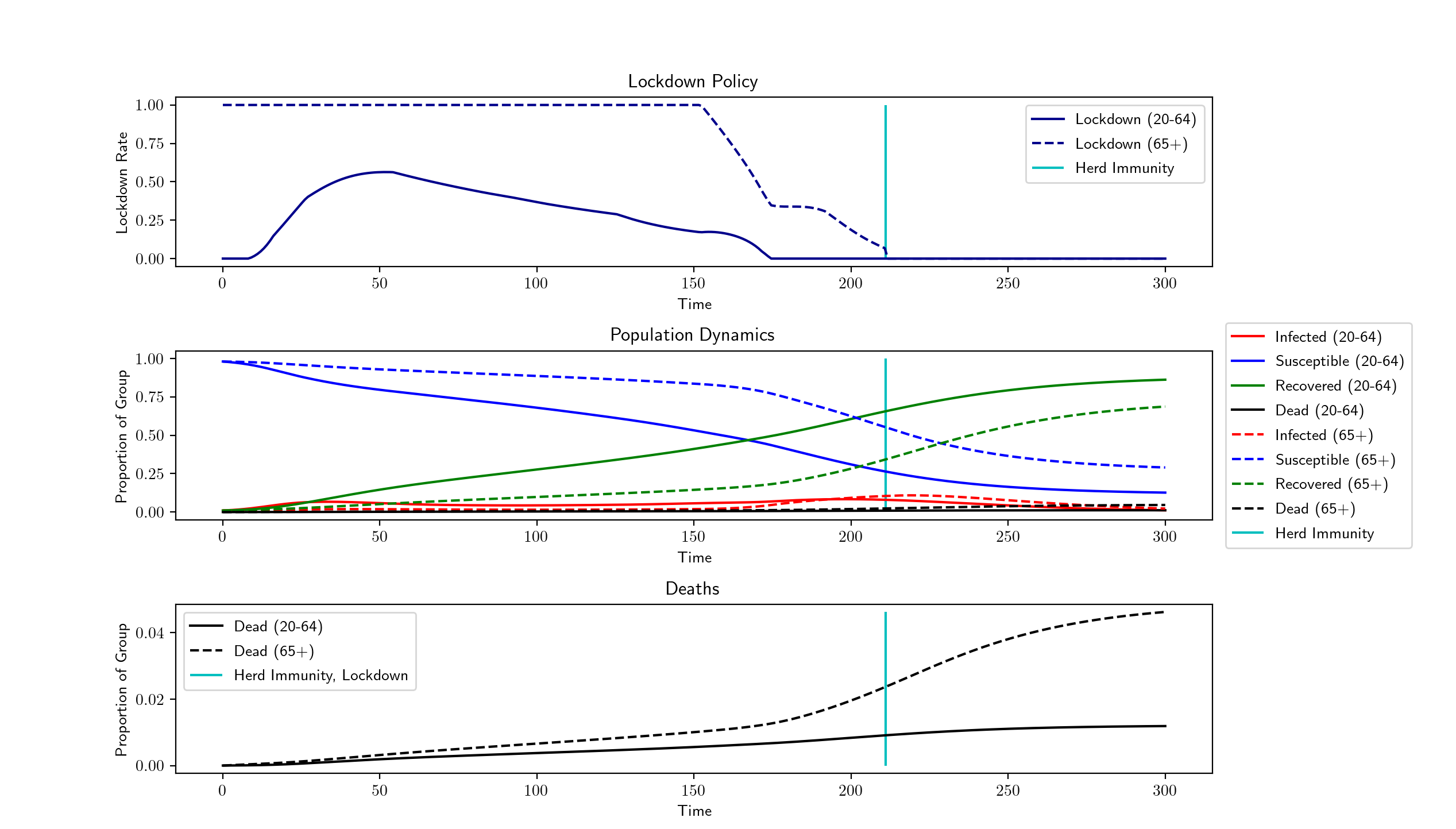}
    \caption{Our model (two groups, herd immunity), parameters from Table \ref{tab:param} \\Herd Immunity: 211 days, Output Loss: 7.8767\%, Total Deaths: 1.8873\%}
    \label{fig:AAComp}
\end{figure}

Now, we investigate the results of our new model using comparable parameter levels. We incorporate herd immunity, deaths indirectly due to lockdown, ability to work remotely, and behavior-dependent transmission rates. Additionally, we consider the possibility of lost employment after the end of the pandemic, as well as the costs of missed health screenings and a monetary penalty for exceeding ICU capacity. To allow comparisons with previous works, we use death rates of a similar magnitude to those in \cite{alv} and \cite{acem}, but we change some parameters to better fit the current situation. Interest rates have dropped significantly, so we use a 0.001\% interest rate, instead of the 5\% used by \cite{alv} or the 1\% used by \cite{acem}. Note that since the interest rate is extremely low, there is little to no discounting applied to wages. We also lengthen the projected average career length in the low-risk group. Finally, we adjust the population distribution slightly from 21\% high-risk to 18\% high-risk, based on data from the 2010 United States Census \cite{census}. The results of our model using the parameters listed in Table \ref{tab:param} is shown in Figure \ref{fig:AAComp}. Most noticeably, lockdown rates for \textit{both} groups fall to 0 after the entire population reaches herd immunity, which is explicitly imposed by our model. Additionally, the lockdown for the low-risk group is slightly shorter but more intense. Unsurprisingly, incorporating deaths of despair increases the total number of deaths due to the epidemic, but this effect is kept small by the shorter lockdowns. Long-term costs of lockdown (an additional penalty for ICU overcrowding, future deaths due to current health negligence, and future unemployment beyond the lockdown) increase output loss while not directly contributing to deaths during lockdown. However, these output losses are offset by the proportion of the population that is able to work remotely from home and the shorter lockdown periods. 

\subsection{Realistic Death Rates}
\label{sec:realistic}
\begin{figure}
    \centering
    \includegraphics[width = 0.95\textwidth]{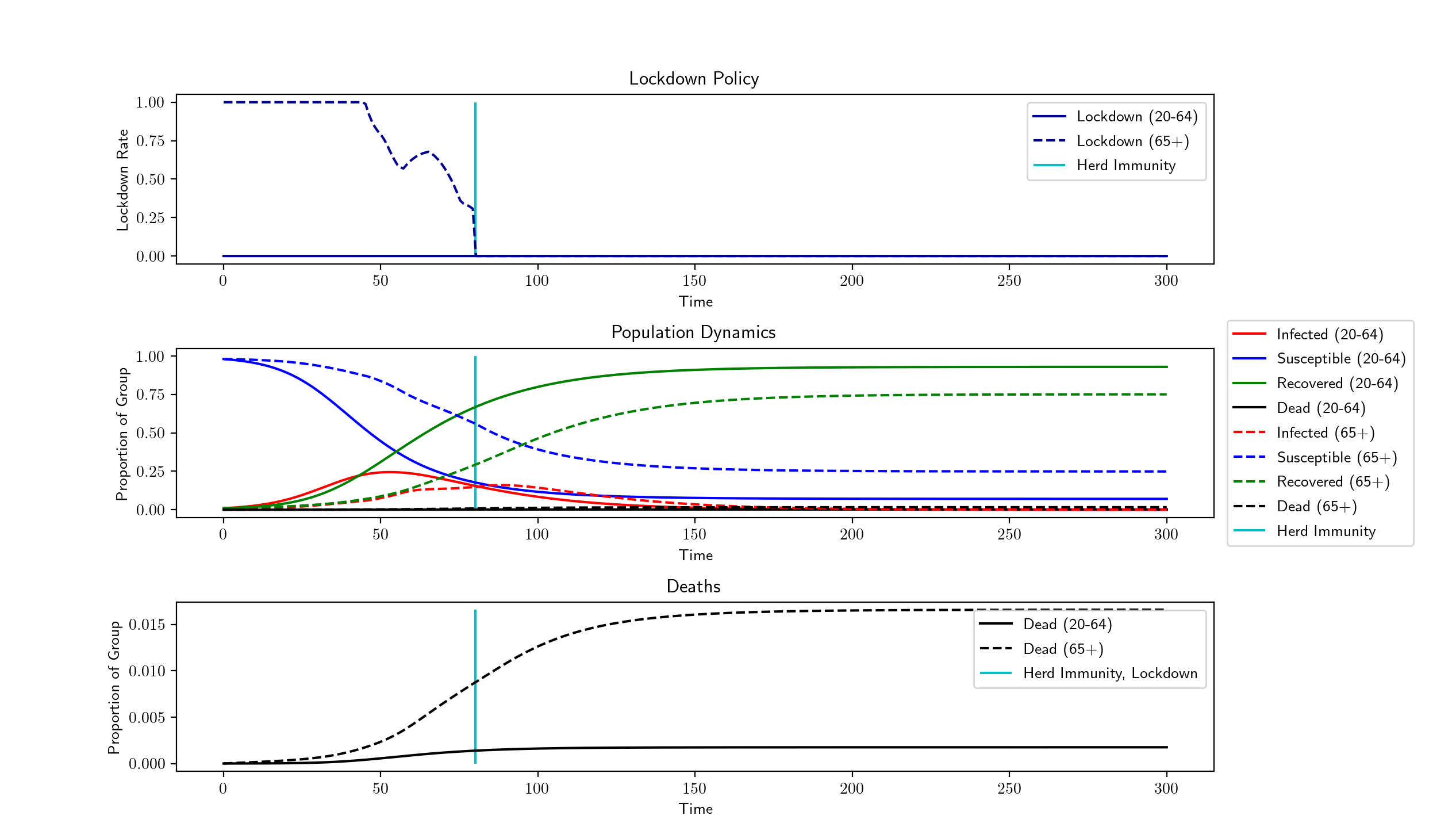}
    \caption{Our model ($\chi = 0.2/r$, death rates from Table \ref{tab:low-death}, all others from Table \ref{tab:param}) \\Herd Immunity: 80 days, Output Loss: 0\%, Total Deaths: 0.4464\%}
    \label{fig:low-chi}
\end{figure}

\begin{figure}
    \centering
    \includegraphics[width = \textwidth]{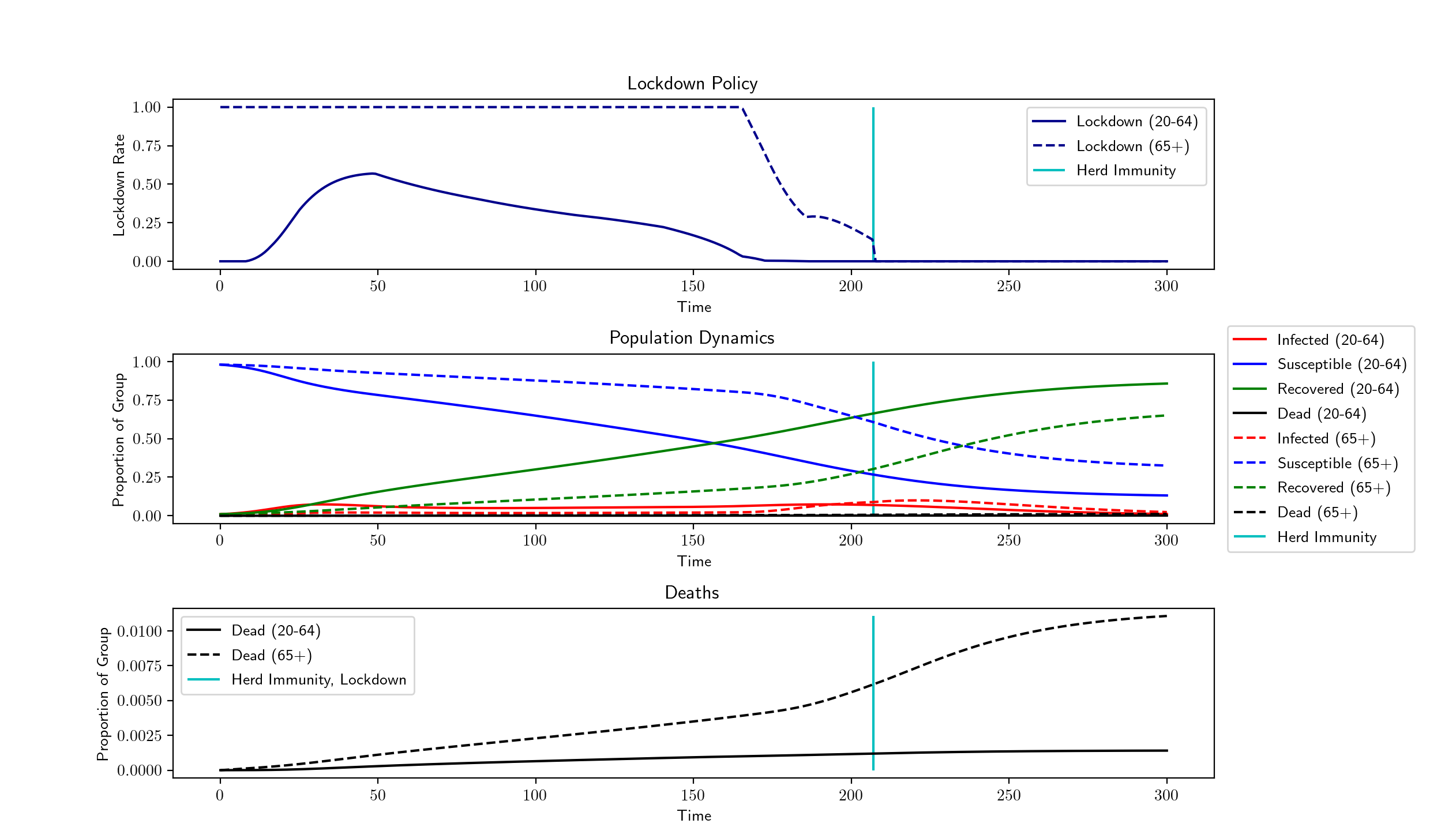}
    \caption{Benchmark -- Herd Immunity: 207 days, Output Loss: 7.3439\%, Total Deaths: 0.3266\% \\
        Benchmark Parameters: $\chi = 10/r$, $r = 0.001\%$,  $\alpha_E = 0.42$, $h$ = 0.4, $\alpha_L = 10^{-5}$, $\alpha_I = 1$, $\rho = 0.75$, F = 1, $\theta = 0.75$, $\nu  = 0.67$, $\eta = 10$,  Death rates from Table \ref{tab:low-death}, Herd Immunity = 60\% }
    \label{tab:robust-table}
    \label{fig:benchmark}
\end{figure}

\begin{figure}
   \centering
    \includegraphics[width = 0.9\textwidth]{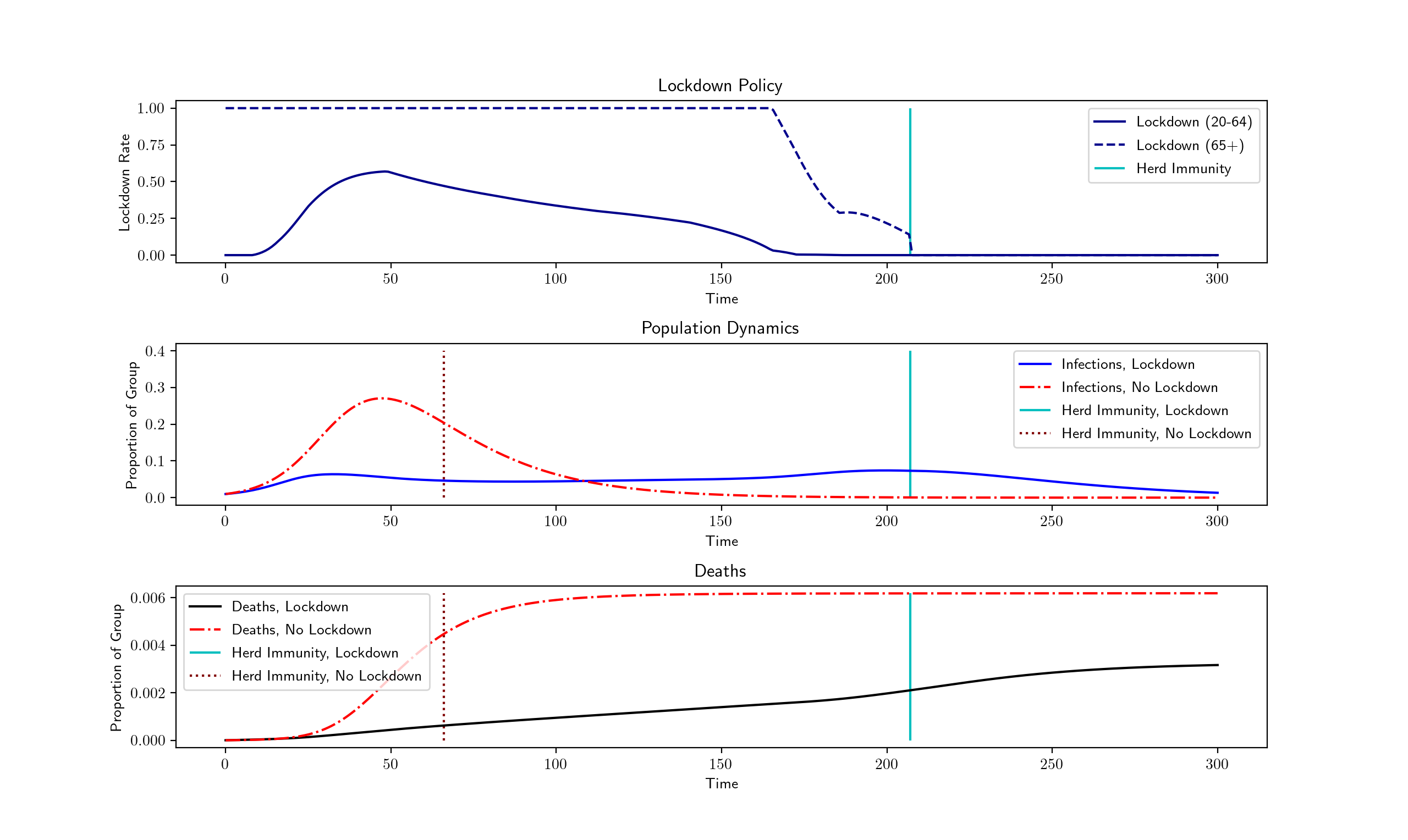}
   \caption{Comparison of optimal lockdown policy to no lockdown, using benchmark parameters, Optimal Lockdown Deaths: 0.3266\% vs Uncontrolled Deaths: 0.6189\%}
   \label{fig:comp}
\end{figure}

Recent CDC reports \cite{CDCstats} indicate that the death rates are much lower than those used in Subsection \ref{sec:calib}. To increase the realism of our model, we update the model death rates according to this newer data\footnote{We use the data in \cite{CDCstats} Table 1 (Scenario 5: Best Current Estimates). To calculate $\delta^0_1$, we construct a weighted average of the Symptomatic Case Fatality Ratio for 0-49 year olds and for 50-64 year olds using 2010 Census data \cite{census} and  multiply by 0.65, since the CDC estimates that 35\% of cases are asymptomatic. For the same reason, we multiply the Symptomatic Case Fatality Ratio for the 65+ group by 0.65 to determine $\delta^0_2$. We set $\delta^1_j$ such that a 30\% infection level causes a five-fold increase in deaths, as in \cite{acem}.} and use them for all subsequent results. These death rates are listed in Table \ref{tab:low-death} and the result, shown in Figure \ref{fig:low-chi}, predicts total mortality of 0.4464\% and total output loss of 0.0013\%. The negligible output loss is due to the negligible lockdown for the low-risk group. However, lockdowns have already been imposed for both groups (and indeed we might desire a death rate lower than 0.4464\%), so we increase the non-pecuniary value of life, $\chi$, and observe how the model changes. By increasing $\chi$ from $0.2/r$ to $10/r$, Figure \ref{fig:benchmark} shows that both groups experience levels of lockdown similar to that of Figure \ref{fig:AAComp}, but with an output loss of 4.8984\% and a lower total death toll of only 0.3266\%. We designate this the \textit{benchmark} situation, which uses death rates from Table \ref{tab:low-death} and $\chi = 10/r$ but keeps all other parameter values consistent with those in Table \ref{tab:param}. We also compare the results of the optimal lockdown to those generated by an uncontrolled scenario with the same parameters, shown in Figure \ref{fig:comp} and Table \ref{tab:comp2}. Without lockdown there is no output loss, but final mortality numbers are approximately twice as high. 

\subsection{Varying Initial Conditions}
\begin{figure}
    \centering
        \begin{subfigure}{\textwidth}    
               \includegraphics[width = \textwidth]{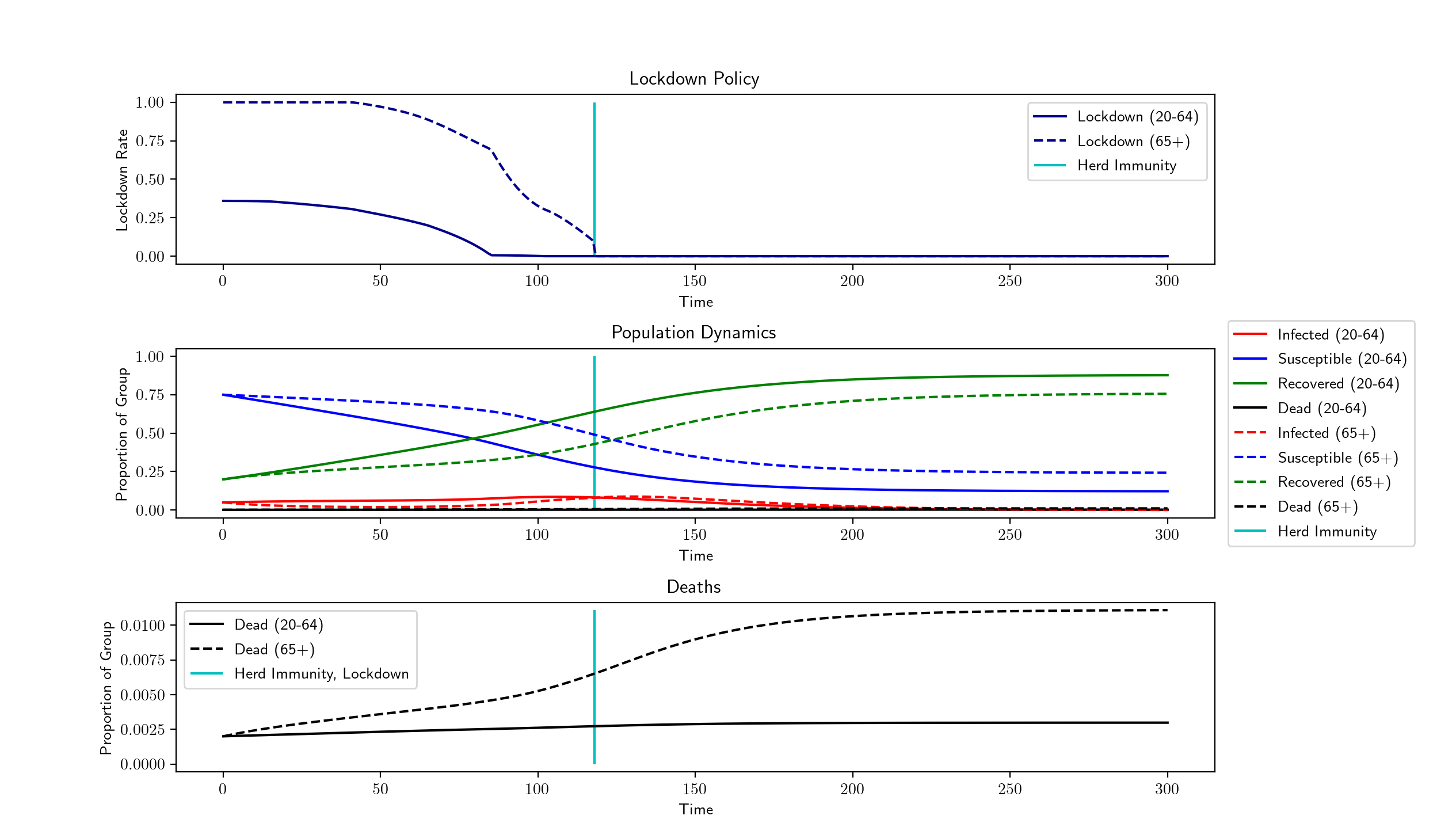}
               \caption{$S_0 = 74.8\%,\ I_0 = 5\%\, R_0 = 20\%,\ D_0 = 0.2\%$, Lockdown: 118 days, Output Loss: 3.4677\%, Additional Deaths: 0.2452\%}
               \label{fig:inita}
       \end{subfigure}
        \begin{subfigure}{\textwidth}    
            \includegraphics[width = \textwidth]{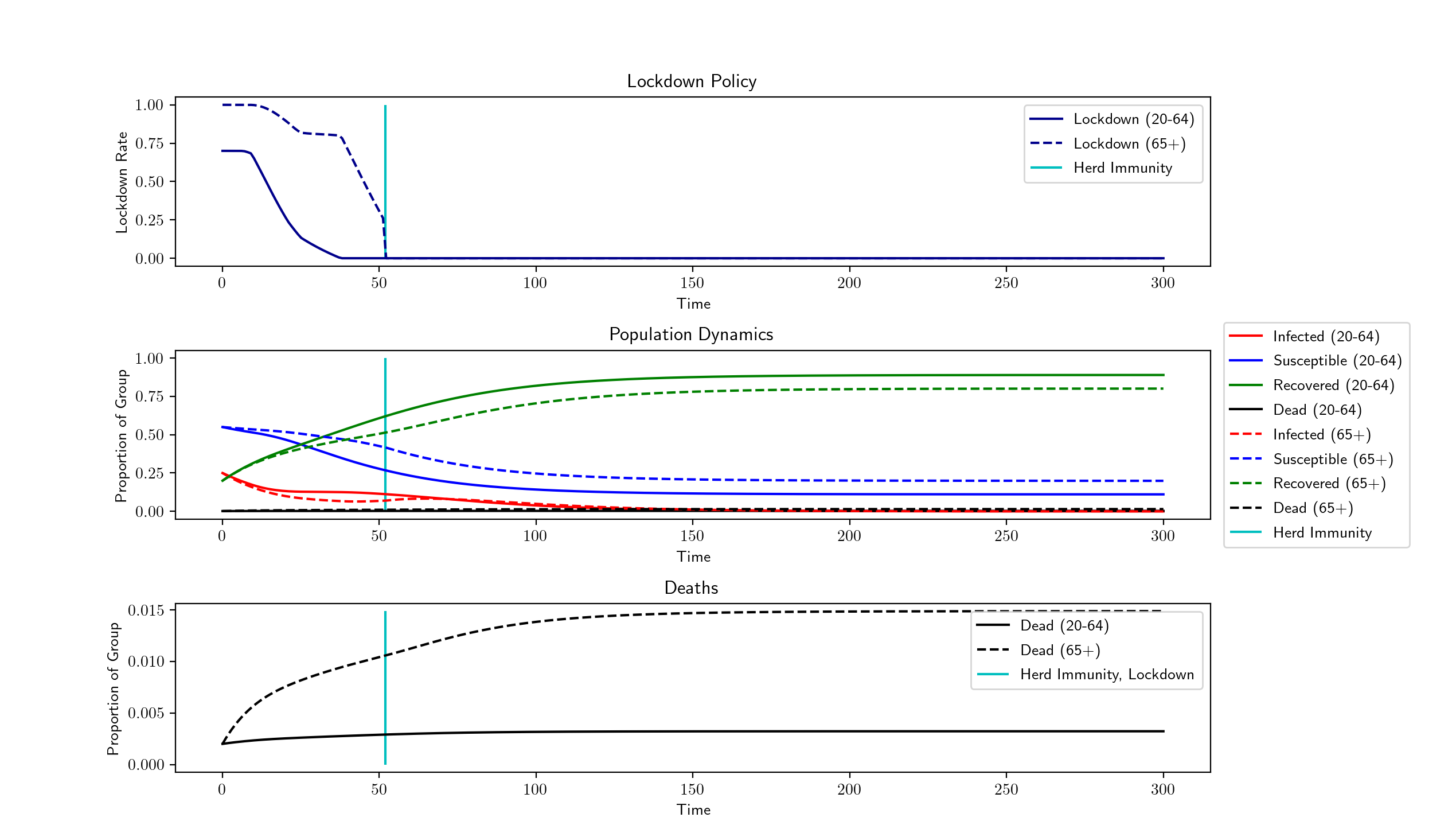}
   	 \caption{$S_0 = 54.8\%,\ I_0 = 25\%\, R_0 = 20\%,\ D_0 = 0.2\%$, Lockdown: 52 days, Output Loss: 2.1755\%, Additional Deaths: 0.3346\%}
	 \label{fig:initb}
        \end{subfigure}
    \caption{Results for varied initial conditions using benchmark parameters}
\end{figure}

Since we are currently in the middle of the pandemic, we investigate at how different initial conditions change the recommended lockdown levels. We model a situation where the pandemic is ongoing and lockdown measures have been lifted, but a sudden spike in infections occurs which prompts new lockdown measures. We consider a case where 20\% of the population has recovered and 0.2\% has died, similar to estimates of the current situation in New York City \cite{nyc}. In Figure \ref{fig:inita}, a small infection spike affects 5\% of the population before lockdown measure are put in place. In this case, we see additional deaths of 0.2452\%. In Figure \ref{fig:initb}, a large infection spike affects 25\% of the population, causing 0.3346\% additional deaths. Note that the lockdown is actually shorter for larger infection spikes, since the larger infection level (which occurs before lockdowns are imposed) moves the population closer to herd immunity. The price of this shorter lockdown, though, is higher mortality rates. 

\subsection{Parameter Robustness}

It is natural to ask how changes in other parameters affect the optimal controls. In general, changes in parameters create the expected changes in lockdown length and intensity, output loss, and mortality. The more interesting question asks about the \textit{level} of impact of various parameters. The effects of non-pecuniary value of life ($\chi$) have already been discussed and are displayed in Figures \ref{fig:low-chi} and \ref{fig:benchmark} and Table \ref{tab:death-comp}, but now we discuss other parameters. In Table \ref{tab:robust-table}, we list the lockdown levels, mortality, and output loss for various other configurations of parameter choices. 

There are some elements of the model which do not have large impacts on the results. The ICU overcapacity constraint barely affects results, likely due to a combination of low infection rates and a sufficiently high number of ICU beds (on average) in the United States. This allows ICU admittance rates to remain at or below the threshold. The expected vaccine arrival date also does not have much effect on the optimal lockdown levels, since the population is expected to reach herd immunity well before its introduction. Based on the current situation, it seems extremely unlikely that a vaccine will be developed less than 8 months after the start of the outbreak, so we considered expected vaccine arrival times of 1 year ($\nu = 1$) and 8 months ($\nu = 1.5$) and found that neither adjustment had much effect. It can also be seen from the table that removing $F$, the cost representing future deaths due to current lack of health maintenance, lengthens the lockdown only slightly, and has little effect on output loss and mortality. 

\begin{figure}
   \centering
    \begin{subfigure}{\textwidth}
           \includegraphics[width = \textwidth]{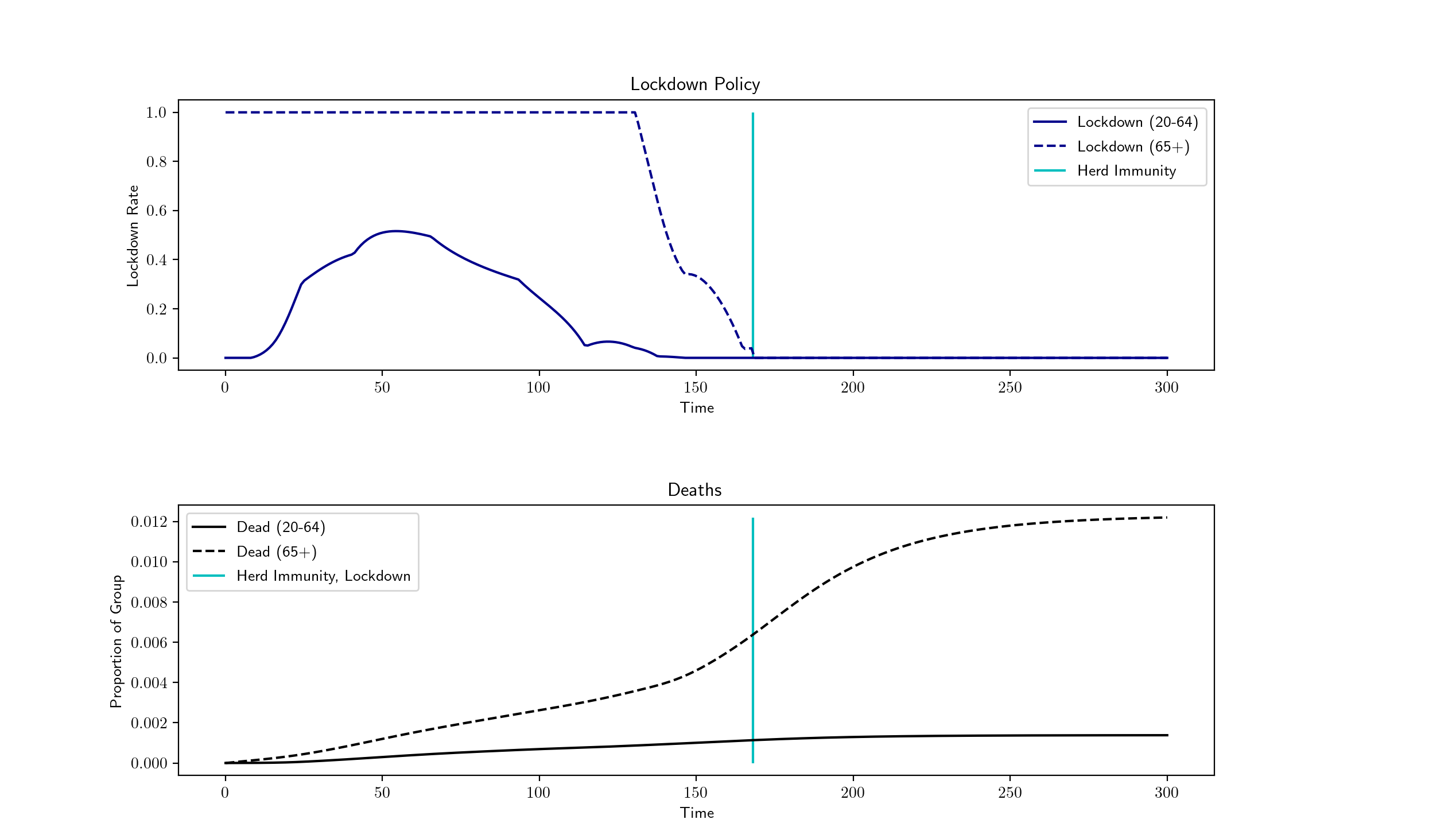}
            \caption{$h = 0$, Lockdown: 168 days, Output Loss = 8.8592\%, Total Deaths =  0.338\%}
       \end{subfigure}
        \begin{subfigure}{\textwidth}
           \includegraphics[width = \textwidth]{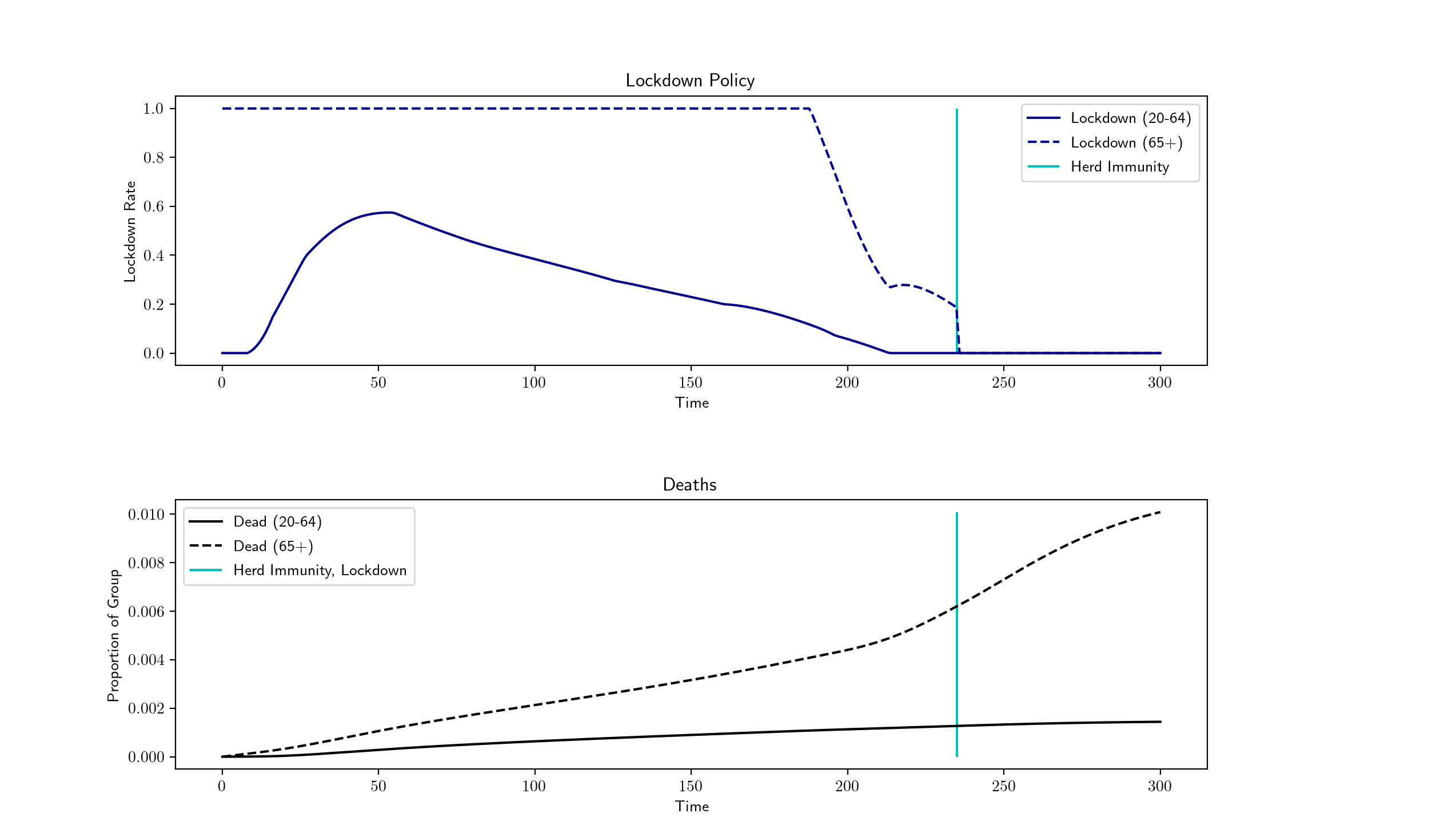}
            \caption{$h = 0.60$, Lockdown: 235 days, Output Loss = 5.842\%, Total Deaths = 0.3251\%}
       \end{subfigure}
    \caption{Robustness results for $h$ (percentage of workforce that can work remotely)\\
    $\rho = 0.75$, $\chi = 10/r$, $r = 0.001\%$, $\nu  = 0.67$, $\alpha_L = 10^{-5}$, $\alpha_I = 1$, $\alpha_E = 0.42$, $\eta = 10$, F = 1}
   \label{fig:robust-wfh}
\end{figure}

In contrast, loss of future employment, ability to work remotely, indirect death rate in lockdown, inter-group interaction, and behavior-dependent infection transmission have significant effects on lockdown length and severity. Adding output savings from remote work, future employment loss due to lockdown, and indirect deaths of lockdown affects the wider population and so produce similar changes in outcomes. The model produces uniformly better outcomes when the level of remote work, $h$, is increased, though lockdowns do last longer as seen in Figure \ref{fig:robust-wfh}. Intuitively, this follows from the idea that more people working remotely helps to maintain economic activity without increasing risk of infection. Changing $\alpha_I$ and $\alpha_L$ in the opposite direction of remote work creates similar effects on the optimal lockdown policy, though outcomes move differently. Increasing the length of projected future unemployment, $\alpha_E$, leads to a shorter lockdown and less output loss, while deaths increase.  This suggests that $\alpha_E$ influences the trade-off between output loss and mortality. And, when we look at varying values of $\alpha_L$, we can see what happens when the optimization tries to minimize deaths due to COVID-19 while also trying to avoid deaths due to lockdown. When $\alpha_L=0$ and the model doesn't take indirect deaths into account, the lockdown extends for longer and has a larger effect on output, but we see much lower mortality levels. However, when $\alpha_L$ is increased to 50 deaths per 100,000 individuals at full lockdown, the model dramatically shortens the lockdown, which decreases output loss and indirect deaths but which leads to higher deaths due to COVID-19.

\begin{figure}
   \centering
    \begin{subfigure}{0.7\textwidth}
           \includegraphics[width = \textwidth]{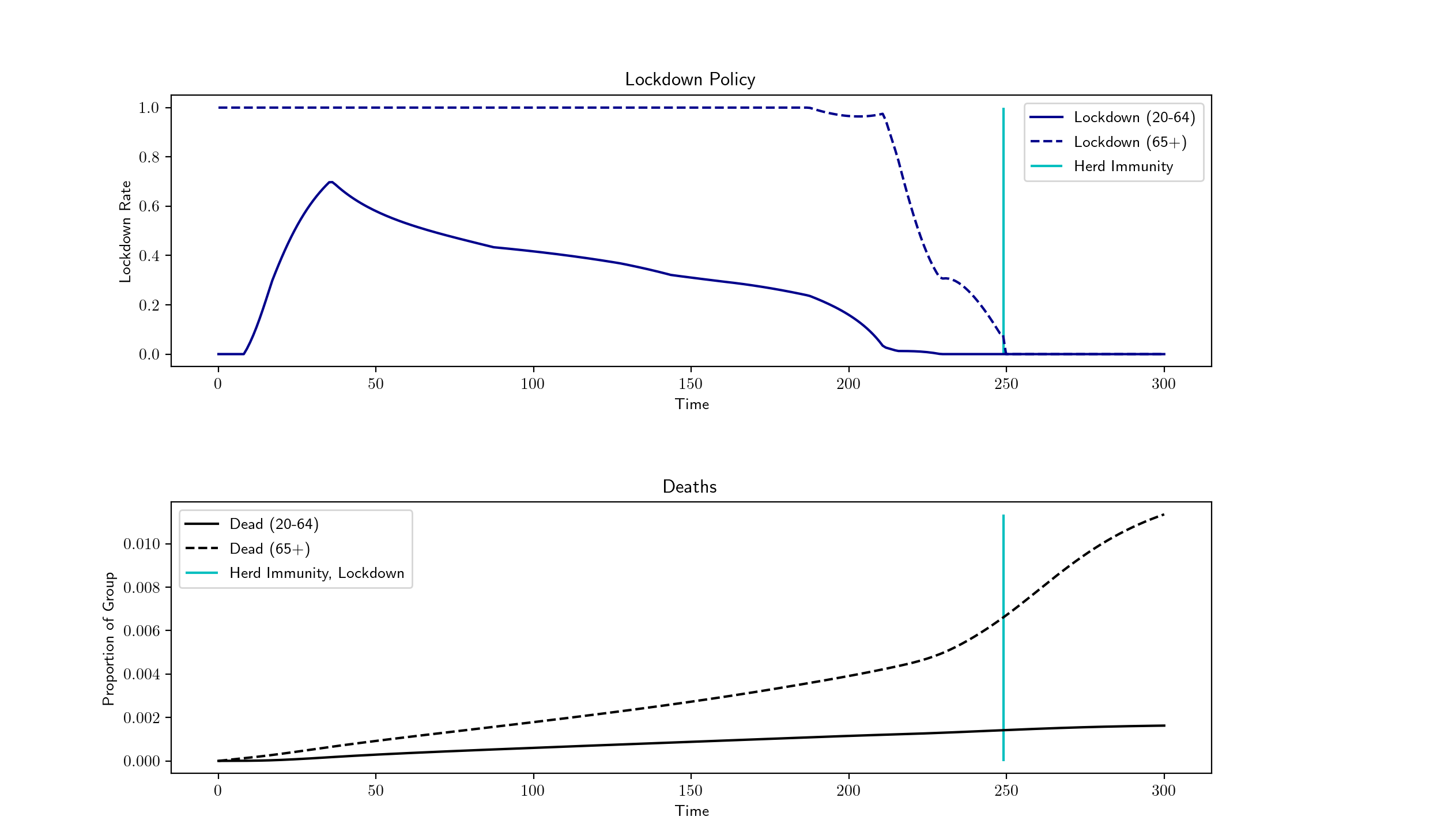}
            \caption{$\alpha_I = 0$, Lockdown: 249 days, Output Loss: 10.6699\%, Total Deaths: 0.3712\%}
       \end{subfigure}
           \begin{subfigure}{0.45\textwidth}
           \includegraphics[width = \textwidth]{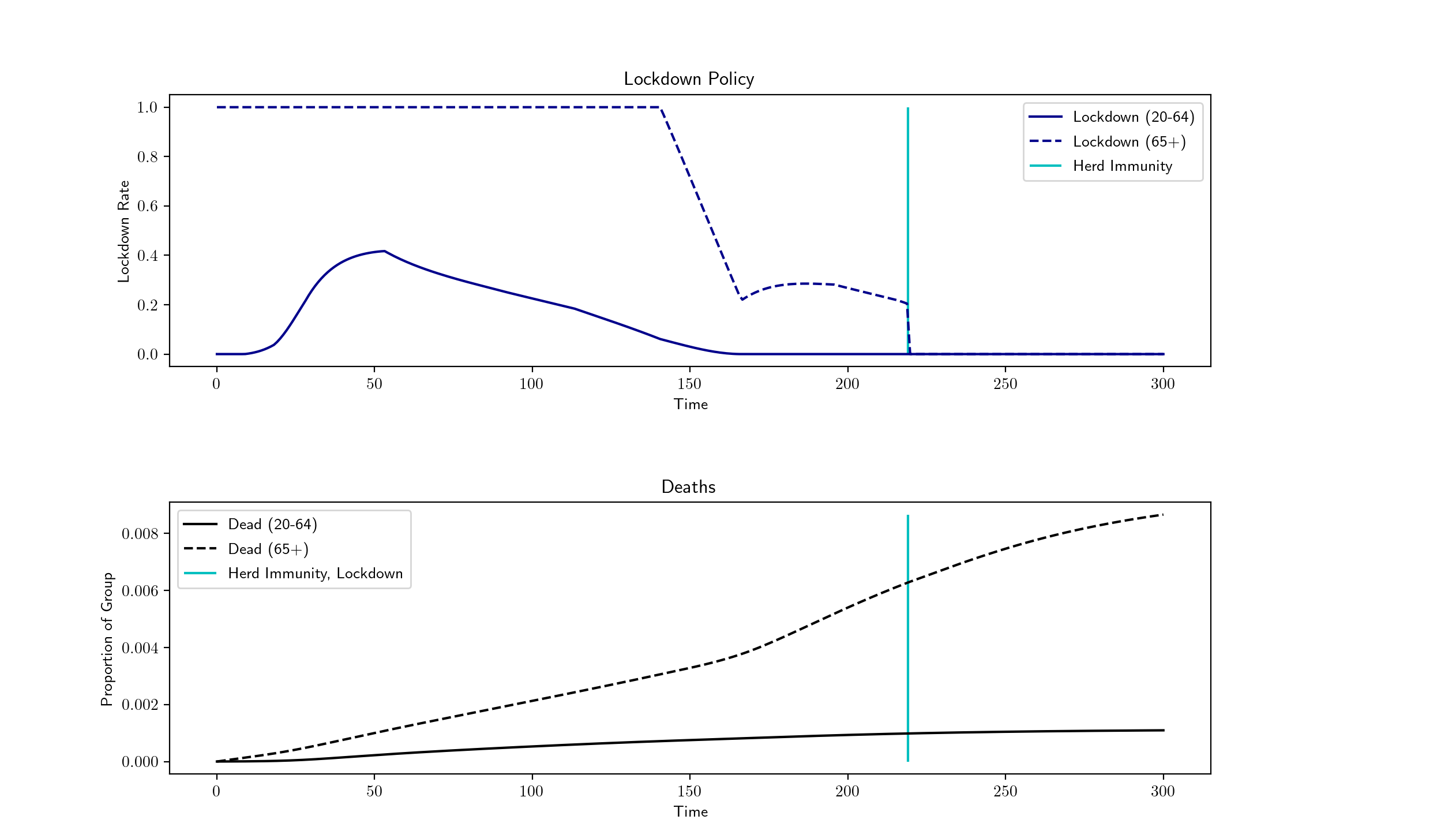}
            \caption{$\alpha_I = 6$, Lockdown: 219 days, Output Loss: 4.5236\%, Total Deaths: 0.2651\%}
       \end{subfigure}
            \begin{subfigure}{0.45\textwidth}
           \includegraphics[width = \textwidth]{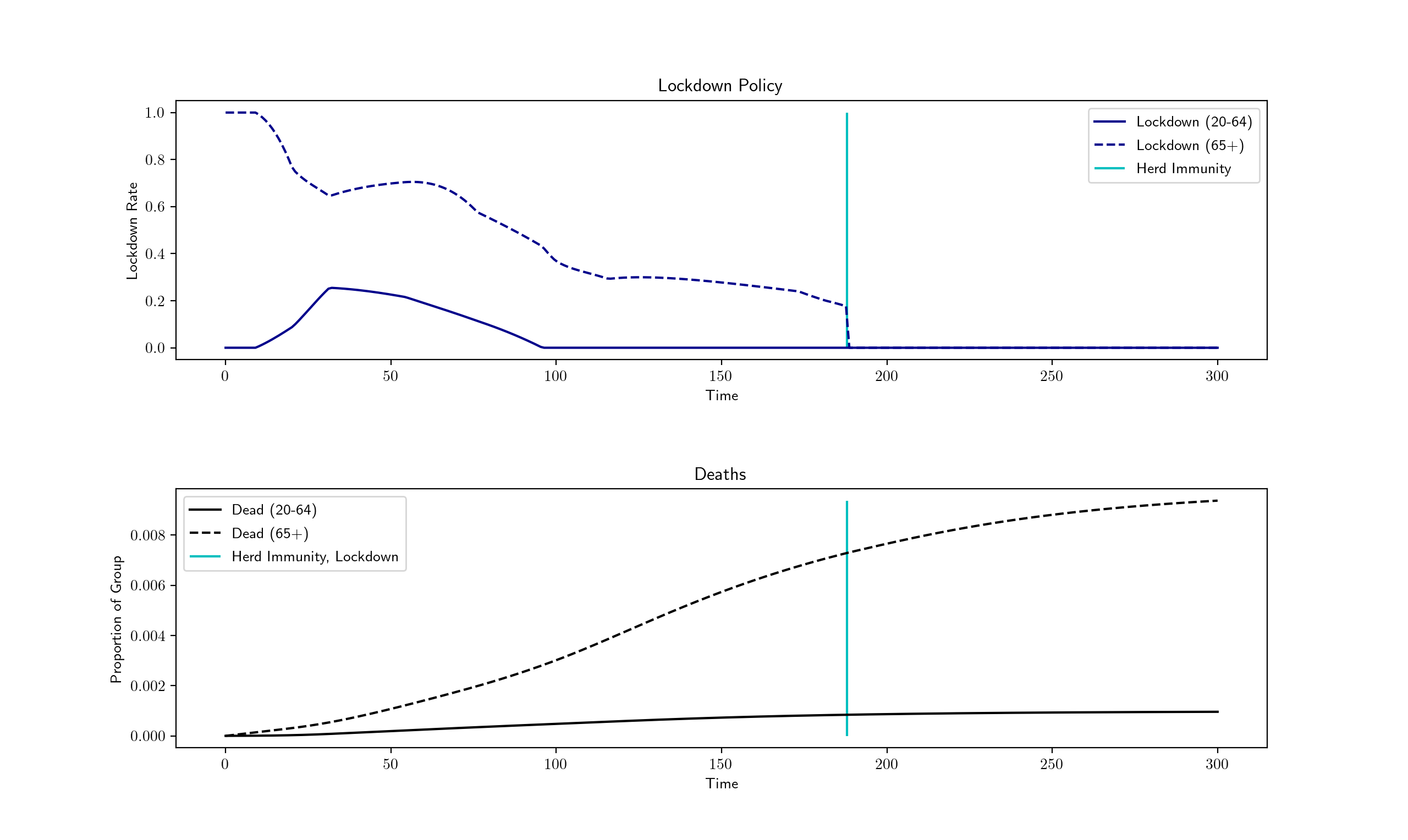}
            \caption{$\alpha_I = 8$, Lockdown: 188 days, Output Loss: 1.9019\%, Total Deaths: 0.259\%}
       \end{subfigure}
        \begin{subfigure}{0.7\textwidth}
           \includegraphics[width = \textwidth]{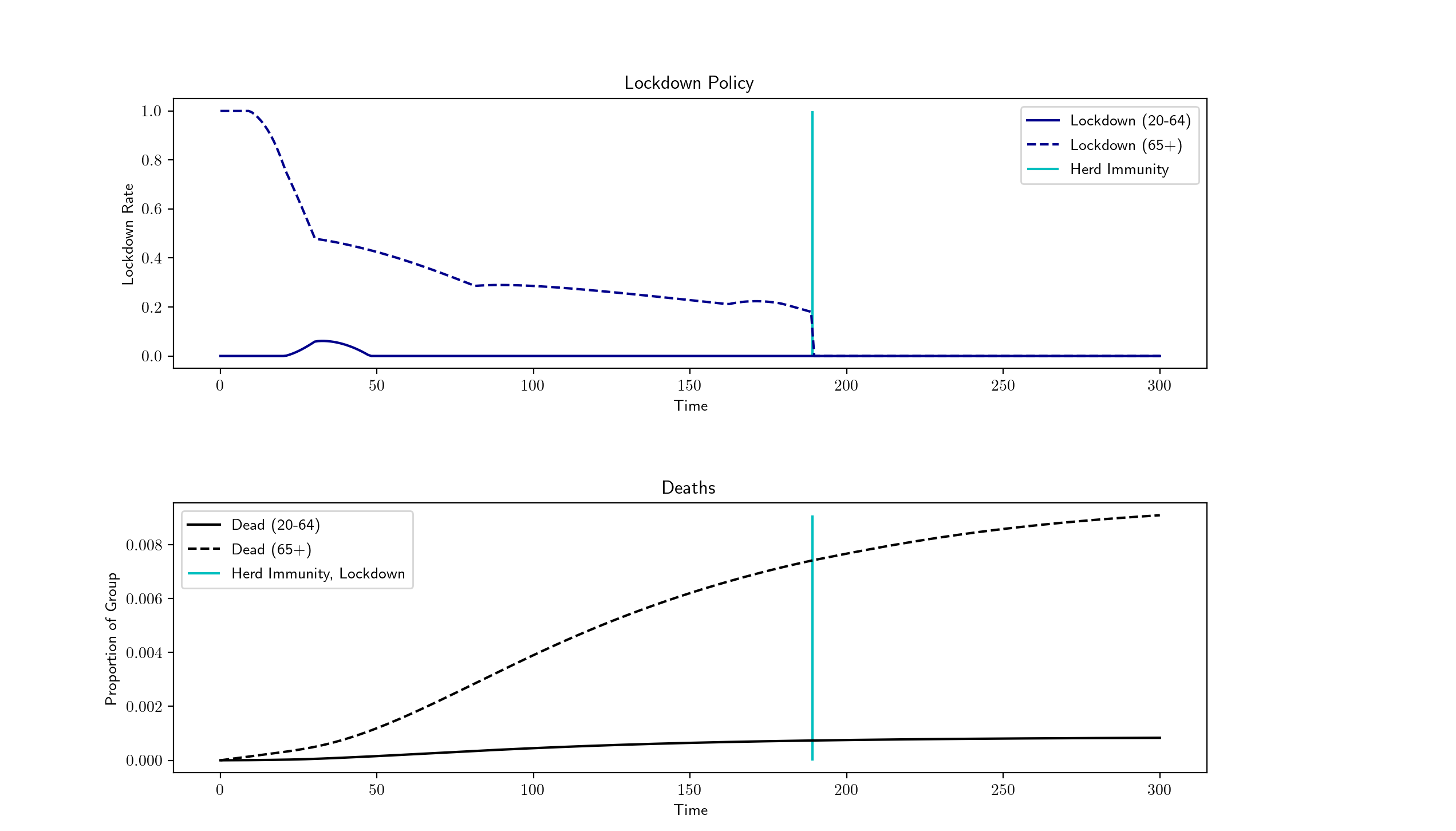}
            \caption{$\alpha_I = 10$, Lockdown: 189 days, Output Loss = 0.1558\%, Total Deaths = 0.245\%}
       \end{subfigure}
    \caption{Robustness results for $\alpha_I$ (scale factor for individual carefulness in response to current levels of infection) \\
    $\rho = 0.75$, $\chi = 10/r$, $r = 0.001\%$, $\nu  = 0.67$, $\alpha_L = 10^{-5}$, $\alpha_E = 0.42$, $\eta = 10$, F = 1, $h = 0.4$}
   \label{fig:robust-bsir}
\end{figure}

\begin{figure}
   \centering
    \begin{subfigure}{\textwidth} 
           \includegraphics[width = \textwidth]{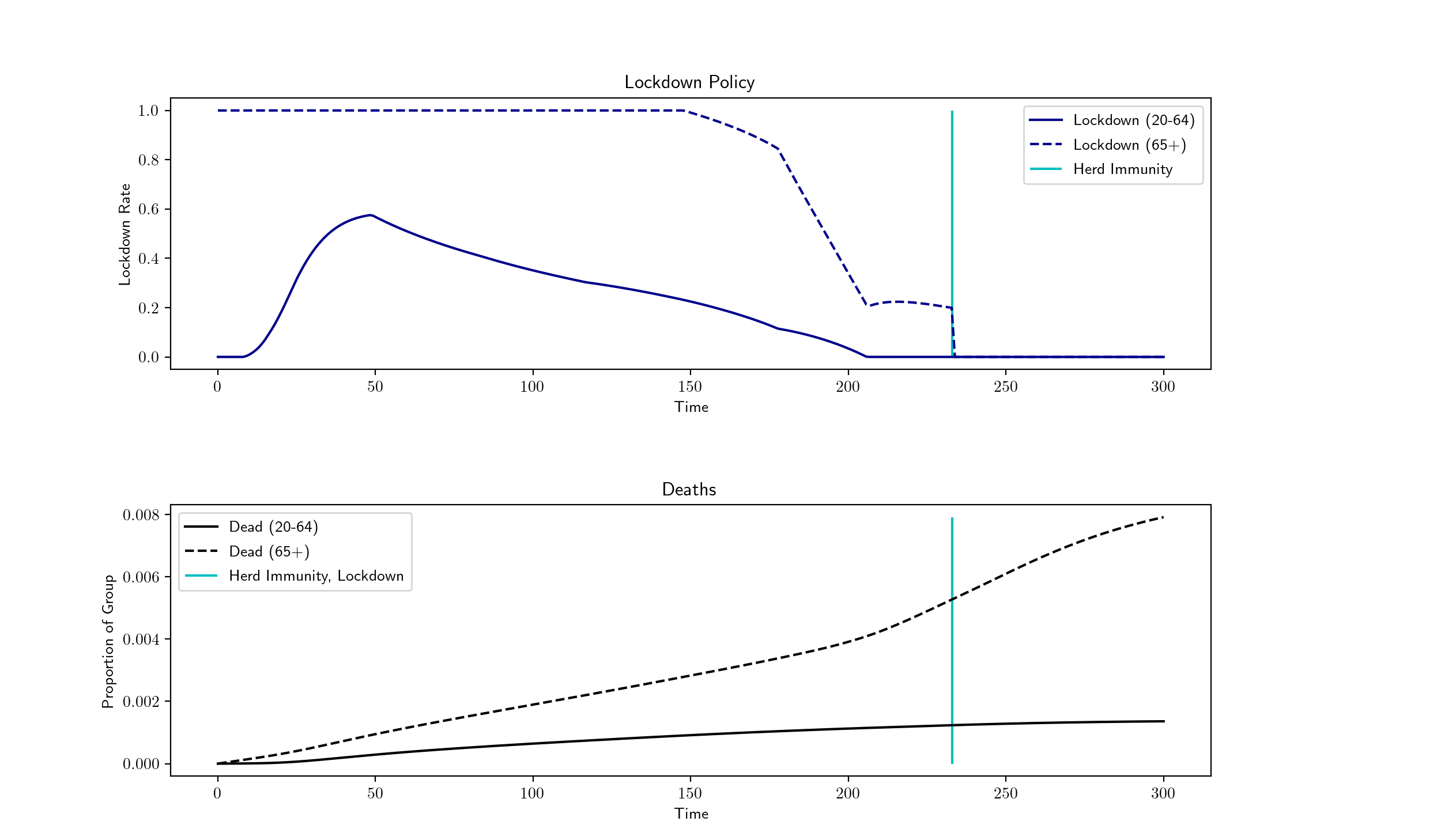}
            \caption{$\rho = 0.5$, Lockdown: 233 days, Output Loss = 8.118\%, Total Deaths = 0.2753\%}
       \end{subfigure}
        \begin{subfigure}{\textwidth}
           \includegraphics[width = \textwidth]{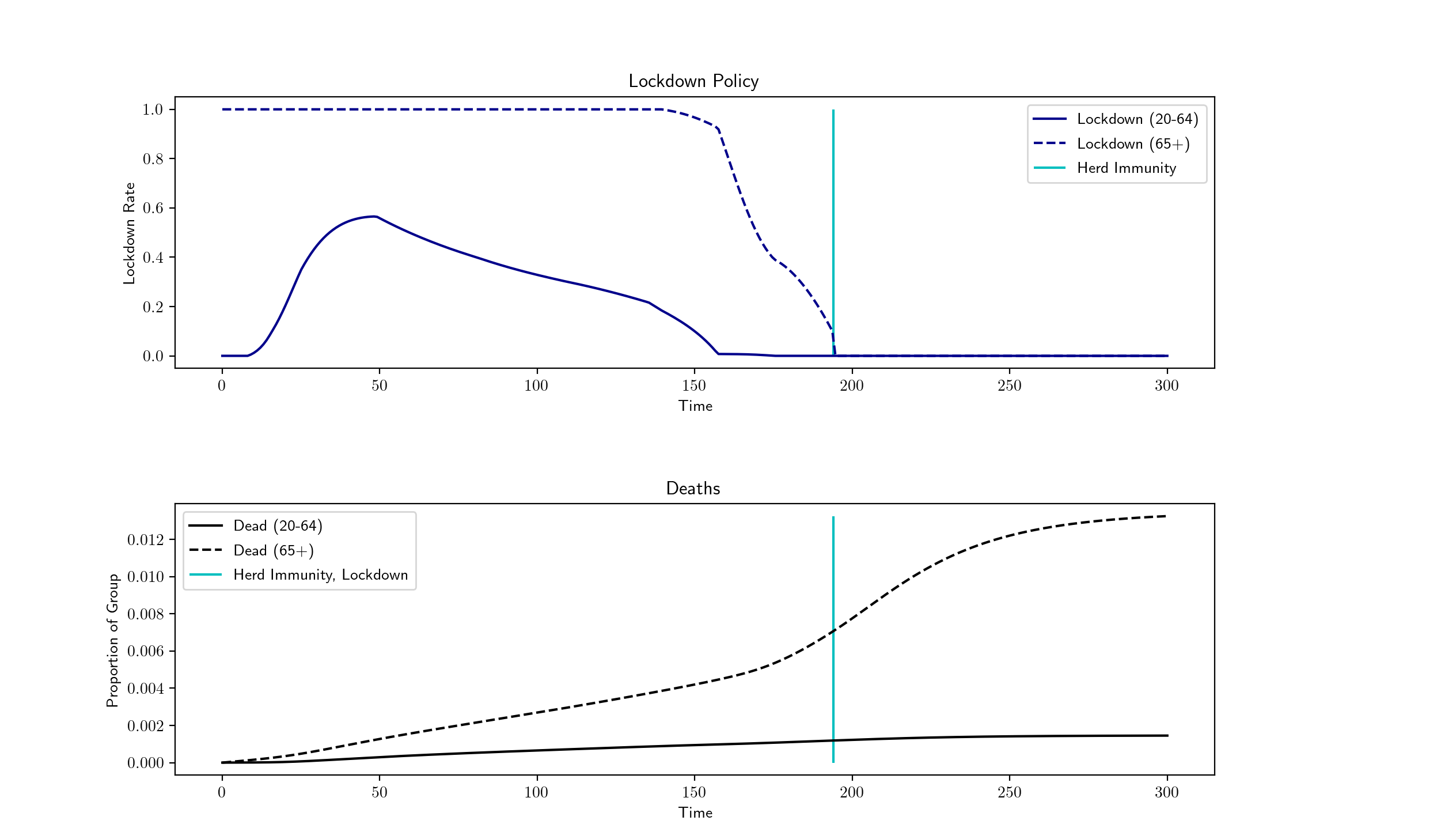}
            \caption{$\rho = 1$, Lockdown: 194 days, Output Loss = 7.0292\%, Total Deaths = 0.3657\%}
       \end{subfigure}
    \caption{Robustness results for $\rho$ (inter-group interaction level) \\
   $\chi = 10/r$, $r = 0.001\%$, $\nu  = 0.67$, $\alpha_L = 10^{-5}$, $\alpha_I = 1$, $\alpha_E = 0.42$, $\eta = 10$, F = 1, $h = 0.4$}
   \label{fig:robust-rho}
\end{figure}

We now discuss the effects of behavior-related parameters. Interestingly, these are the only parameters we discuss which have also an effect on the uncontrolled outcomes. If the level of interaction between groups, $\rho$, is lowered, there is less interaction between the low-risk and high-risk groups and so lockdowns are more effective. In Figure \ref{fig:robust-rho}, when $\rho$ goes from  0.75 to 0.5, the low-risk group is able to begin easing the lockdown earlier since there is less worry about transmission to high-risk individuals. However, the lockdown lasts longer overall, since it is harder to reach herd immunity. This increase in output is offset by a drop in mortality. With the optimal lockdown, mortality decreases to 0.2552\% compared to the benchmark of 0.3166\% when $\rho = 0.75$. With no lockdown, mortality decreases to 0.5268\% from 0.6189\%. The opposite occurs when $\rho$ is increased to 1, meaning that the groups mix freely. Herd immunity arrives sooner due to increased inter-group transmission, however mortality increases to 0.3599\% with lockdown and to 0.6891\% without it. This suggests that it is beneficial for high-risk individuals to exercise extra caution in their interactions with members of the low-risk group. The other parameter that reflects individual behavior, $\alpha_I$, also has a notable impact on optimal lockdown policies. This parameter determines the efficacy of personal actions taken to slow transmission of COVID-19. In the benchmark case, $\alpha_I = 1$. If transmission rates are constant ($\alpha_I = 0$), then the lockdown lasts longer due to the increased likelihood of transmission and mortality increases to 0.3394\%. In the uncontrolled case, the mortality increases too, to 0.7586\%. But, if $\alpha_I$ is set very high (say 10, which implies that personal caution can reduce the transmission rate by 95\%  when infections reach 30\%), then a lockdown is barely necessary, as shown in Figure \ref{fig:robust-bsir}. In this case, uncontrolled mortality is a mere 0.2581\%. This second scenario is perhaps too optimistic, but it demonstrates the potential power of social distancing. If we examine $\alpha_I = 6$ and $\alpha_I = 8$, we see that the lockdowns increase as $\alpha_I$ decreases. If we consider the more modest change from $\alpha_I = 0$ to $\alpha_I = 1$ and decrease the interaction level between groups from $\rho = 0.75$ to $\rho = 0.5$, then imposing the optimal lockdown decreases overall mortality from 0.7586\% to 0.2552\%.

If the population is less obedient and disregards lockdown measures, for example if $\theta$ decreases from 0.75 to 0.6, we see that the lockdown is shorter because herd immunity is reached sooner. This decreases output loss, but leads to more deaths. If the population is more obedient, however, for example if $\theta = 0.85$, the effect of a given level of lockdown is larger with respect to the same level of output loss so lockdowns are less intense but last longer. This slows the arrival of herd immunity, leading to higher output loss, but has the benefit of lower mortality rates. 

\begin{figure}
   \centering
          \begin{subfigure}{0.7\textwidth}
           \includegraphics[width = \textwidth]{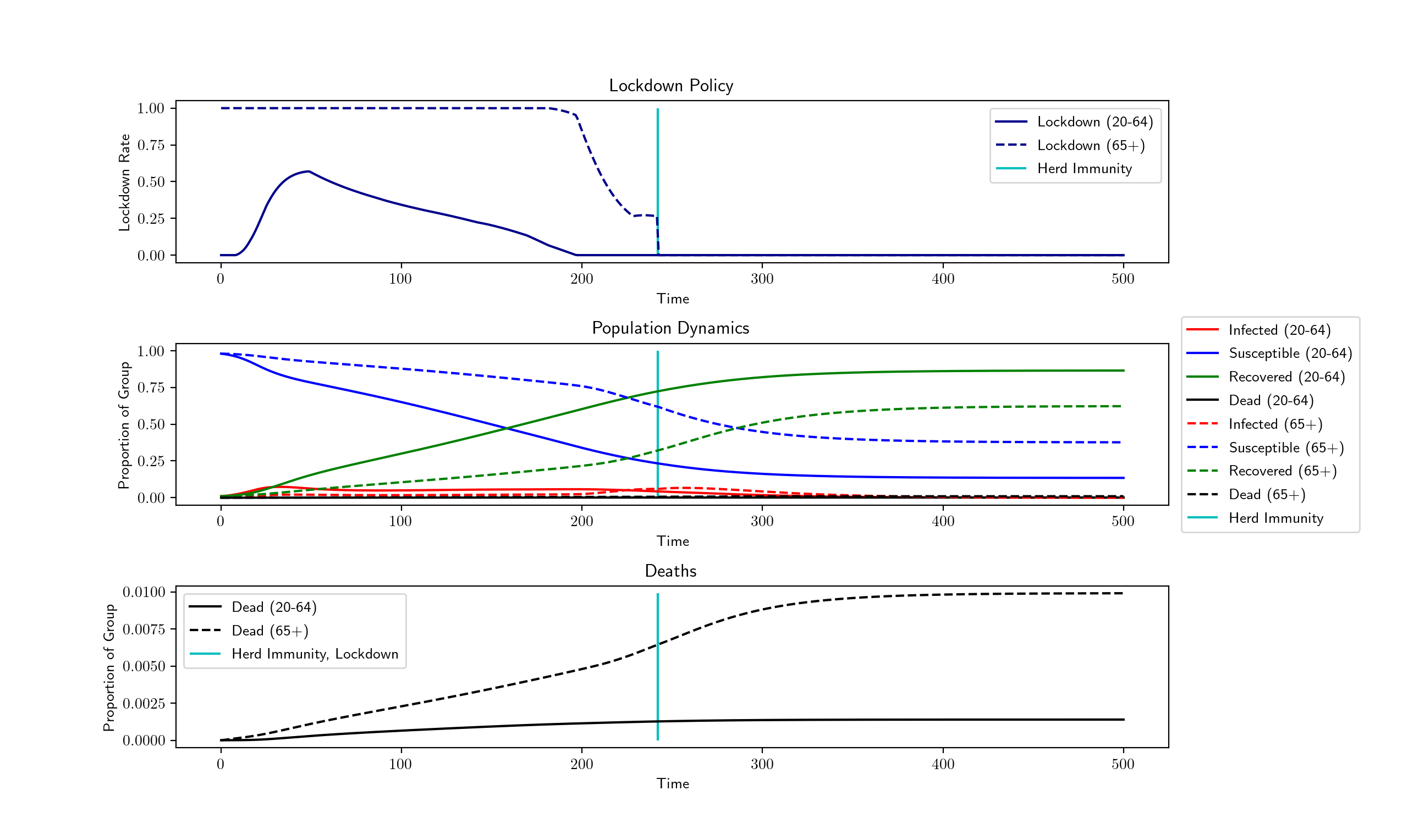}
            \caption{Herd Immunity = 65\%, Output Loss: 7.8084\%, Total Deaths: 0.2948\%}
       \end{subfigure}
          \centering
          \begin{subfigure}{0.7\textwidth}
           \includegraphics[width = \textwidth]{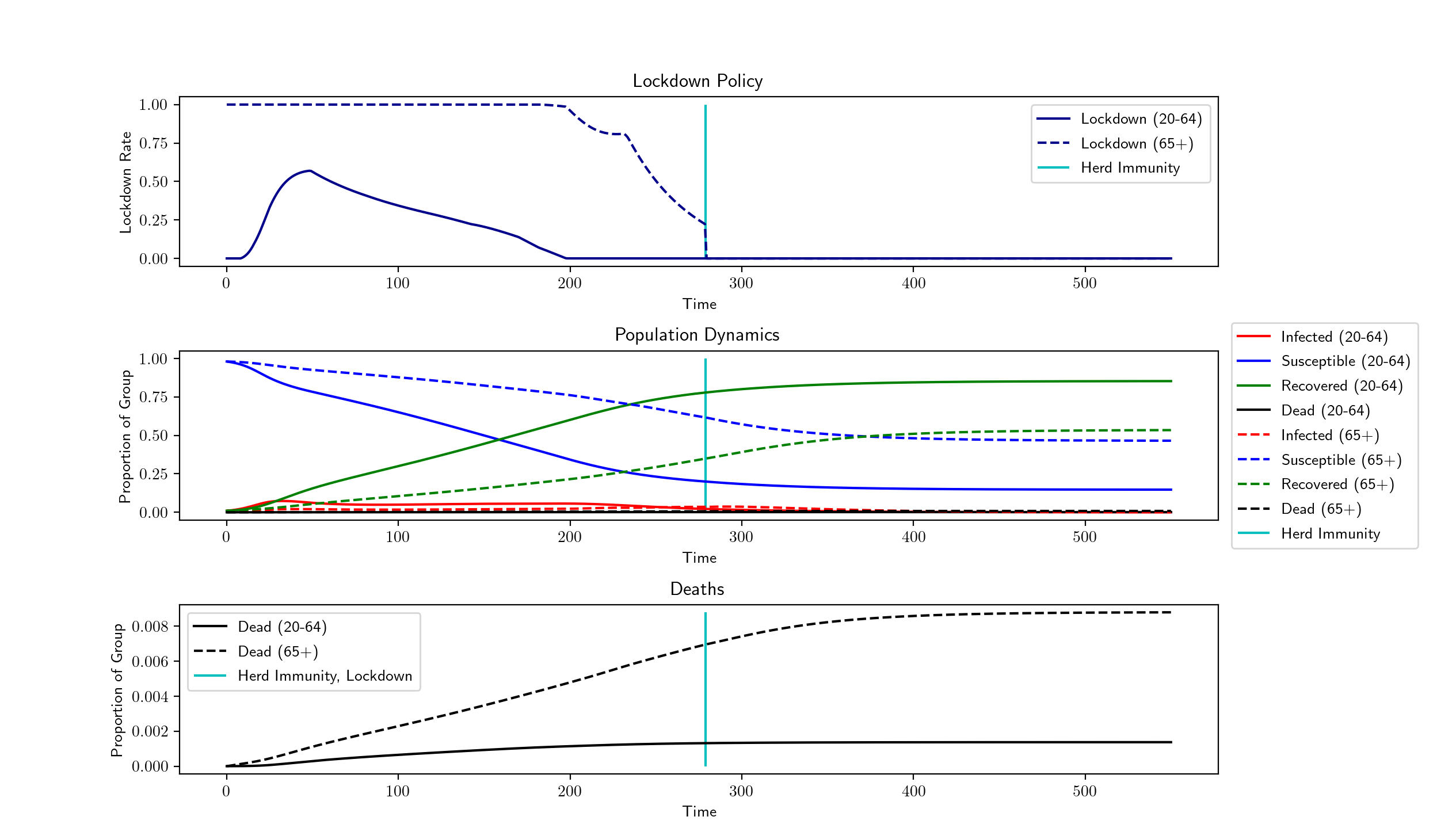}
            \caption{Herd Immunity = 70\%, Output Loss =  7.8496\%, Total Deaths =  0.2724\%}
       \end{subfigure}
          \begin{subfigure}{0.7\textwidth}
           \includegraphics[width = \textwidth]{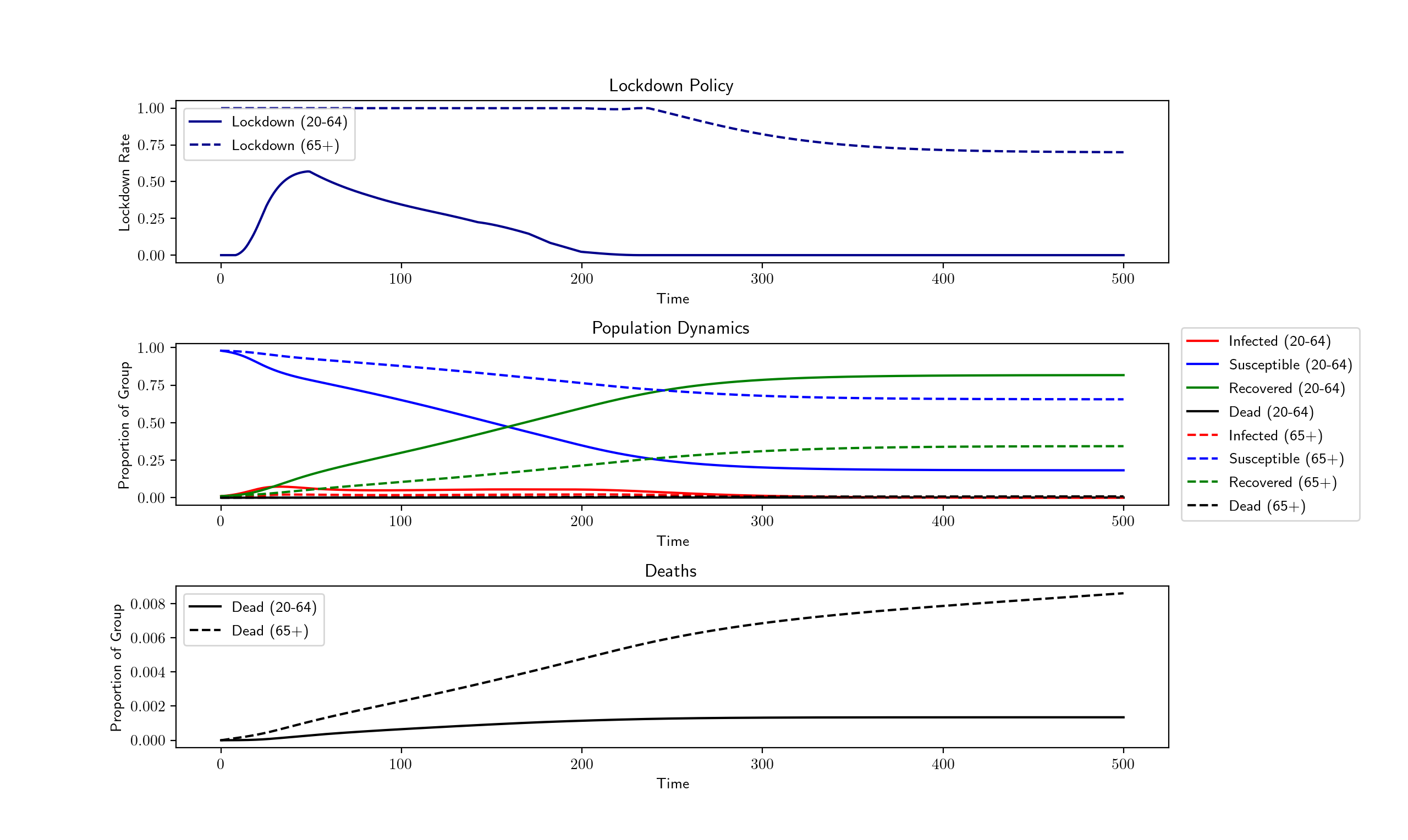}
            \caption{Herd Immunity = 75\%, Output Loss: 7.967\%, Total Deaths: 0.2732\%}
       \end{subfigure}
           \caption{Robustness results for $\sigma$ (arrival of herd immunity)\\
    $\rho = 0.75$, $\chi = 10/r$, $r = 0.001\%$, $\nu  = 0.67$, $\alpha_L = 10^{-5}$, $\alpha_I = 1$, $\alpha_E = 0.42$, $\eta = 10$, F = 1, $h = 0.4$ (cont.)}
   \label{fig:herd}
 \end{figure}

Finally, in Figure \ref{fig:herd} we investigate the effect of increasing the herd immunity threshold $\sigma(x) = \min\{t \geq 0: R_t \geq x\}$. In all of our examples, the low-risk group is able to leave lockdown before the arrival of herd immunity, so moving this threshold does not have much impact on output loss. However, this parameter has important implications for the high-risk group. From Figure \ref{fig:herd}, we see that there is a clear change in dynamics for thresholds of 75\% and above, where the lockdown policy becomes comparable to that of \cite{acem}. The length of lockdown for the low-risk group increases by about 20 days, but the high-risk group remains in lockdown until the vaccine arrives -- an increase of almost 300 days. This abrupt change in strategy arises because eventually the population reaches a steady state with very low infections. The number of susceptible individuals decreases very slowly and is driven largely by deaths due to lockdown, so will not achieve thresholds of 75\% and over before the vaccine is expected to arrive. In this case, the pandemic has been neutralized even though the population has not reached the threshold prescribed by the parameters. These experiments demonstrate that there is not much benefit to be found from a planner overestimating the herd immunity threshold. If the herd immunity threshold is 100\%, this is equivalent to removing the herd immunity exit time. The impact of $\sigma$ can therefore be observed as an decrease in output loss and an increase in mortality, while decreasing lockdown length for the high-risk group by 340 days and for the low-risk group by 45 days.

\section{Appendix - Tables and Figures}

\begin{table}
\bgroup
\def\arraystretch{1.5}%
    \begin{tabular}{|p{0.12\linewidth}|p{0.35\linewidth}|p{0.12\linewidth}|p{0.22\linewidth}|p{0.15\linewidth}|}
        \hline
        \textbf{Parameter} & \textbf{Description} & \textbf{\cite{alv}} & \textbf{\cite{acem}} & \textbf{Our Model} \\
        \hline
        $\bar L$ & Maximum attainable lockdown  & 0.7 & [0.7, 0.7, 1] & [0.7, 1] \\
        \hline
        $\gamma$ & Recovery rate & 1/18 & 1/18 & 1/18 \\
        \hline
        $\delta^0_j$ & Base mortality & 0.01$\gamma$ & $[0.001\gamma, 0.01\gamma, 0.06\gamma]$ & $[0.01\gamma, 0.06\gamma]$ \\
        \hline
        $\delta^1_j$ & Rate of mortality increase based on infection levels & 0.05$\gamma$ & if I = 30\%  then mortality rates are 5 times the base rates &$[0.06\gamma, 0.10\gamma]$ \\
        \hline
        $\iota_j$ & Rate of ICU admittance & N/A & $\sigma$ (unknown) & [0.026, 0.074]\\
        \hline
        $ICU$ & ICU capacity as proportion of overall population (based on beds/100,000 individuals) & N/A & N/A & 0.0003\\
        \hline
        $\eta$ & Scale factor for cost of ICU overcapacity & N/A & N/A & 10 \\
        \hline
        $\beta_0$ & Initial transmission rate & 0.2 & 0.2 & 0.2 \\
        \hline 
        $\rho$ & Interaction level between groups & N/A & 1 & 0.75 \\ 
        \hline
	    $\nu$ & Intensity for vaccine/cure arrival & 0.667/365 (1.5 yrs) &  0.667/365 & 0.667/365\\
	    \hline 
	    $\omega_j$ & Normalized individual daily productivity & 1 & [1, 1, 0] & [1,0] \\
	    \hline 
	    $h$ & Proportion of workforce which can work remotely & 0 & 0 & 0.4 \\
	    \hline
	    $r$ & Yearly interest rate & 5\% & 1\% & 0.001\% \\
    	\hline 
        $\chi$ & Non-pecuniary cost of death & 0 & 20 & 0.2/r \\
        \hline
        $\Delta_j$ & Years left in career & $\infty$ & [15, 7.5, 0] & [20, 0] \\
        \hline
        $\theta_j$ & Obedience of lockdown & 0.5 & 0.75 & 0.75 \\
        \hline
        $ \alpha_L$ & Scaling factor for indirect deaths due to lockdown & 0 & 0 & 0.00001 \\
        \hline
        $\alpha_I$ & Scaling factor for decrease in $\beta_t$ due to personal social distancing measures (masks, etc.) & 0 & 0 & 1 \\
        \hline
        $\alpha_E$ & Scale factor for decreasing potential career length/increasing chance of bankruptcy as lockdown lengthens & N/A & 0 & 0.01 \\
    	\hline
    	$F$ & Future cost of missing health maintenance during lockdown & 0 & 0 & 1 \\
    	\hline
    	$p$ & Flag for immunity passport \newline p=1 $\implies$ no passport \newline p=0 $\implies$ passport & 0 & 0 & 1 \\
    	\hline 
    \end{tabular}
    \caption{Full List of Parameters}
    \label{tab:param}
    }
\end{table}

\begin{table}
    \centering
        \bgroup
\def\arraystretch{1.2} 
    \begin{tabular}{|c|c|c|}
    \hline
        \textbf{Group} & \textbf{$\delta^0$ }& \textbf{$\delta^1$} \\
        \hline
        Age 20-64 & $0.000634\times \gamma$ & $0.00845 \times \gamma$  \\
        \hline
        Age 65+ & $0.00845 \times \gamma$ & $0.1\times\gamma$ \\
        \hline
    \end{tabular}
    \caption{Death rates based on \cite{CDCstats}, $\gamma = 1/18$ is recovery rate as listed in Table \ref{tab:param}}
    \label{tab:low-death}
    }
\end{table}
 
\begin{table}
    \centering
\bgroup
\def\arraystretch{1.5}%
    \begin{tabular}{|p{0.12\linewidth}|p{0.1\linewidth}|p{0.1\linewidth}|p{0.1\linewidth}|p{0.1\linewidth}|p{0.1\linewidth}|p{0.1\linewidth}|p{0.1\linewidth}|p{0.1\linewidth}|}
    \hline
        \textbf{Non-pecuniary Value of Life} & \textbf{Output Loss} & \textbf{Total Deaths (All)} & \textbf{COVID-19 Deaths (All)} & \textbf{Total Deaths (20-64)} & \textbf{COVID-19 Deaths (20-64)} & \textbf{Total Deaths (65+)} & \textbf{COVID-19 Deaths (65+)} \\
         \hline
         $\chi = 0.2/r$ & 0\%  & 0.4464\% & 0.4335 & \%0.1433\% &  0.1433\% & 0.3017\% & 0.2902\%  \\
        \hline
        $\chi = 10/r$ (Benchmark) & 7.3439\% & 0.3266\% & 0.2544\% & 0.1201\% &  0.0855\% & 0.2066\% & 0.1689\% \\
         \hline
    \end{tabular}
    \caption{A comparison of mortality rates for different non-pecuniary value of life, death rates from Table \ref{tab:low-death} }
    \label{tab:death-comp}
     }
\end{table}
 
\begin{table}
    \centering
\bgroup
\def\arraystretch{1.5}%
    \begin{tabular}{|c|c|c|c|}
    \hline
        \textbf{Situation} & \textbf{Output Loss} & \textbf{ Total Deaths} \\
        \hline
        No Lockdown & 0\% & 0.6189\%  \\
         \hline
        Optimal Lockdown & 7.3439\% & 0.3266\%  \\
        \hline
    \end{tabular}
    \caption{Comparison of optimal lockdown policy to no lockdown, using benchmark parameters}
    \label{tab:comp2}
     }
\end{table}

\begin{table}
    \centering
    \bgroup
\def\arraystretch{1.2} 
    \begin{tabular}{|p{0.15\linewidth}|p{0.1\linewidth}|p{0.1\linewidth}|p{0.1\linewidth}|p{0.1\linewidth}|p{0.1\linewidth}|p{0.1\linewidth}|p{0.1\linewidth}|}
    \hline
        \textbf{Parameter Values} & \textbf{Avg. Lockdown (20-64)} & \textbf{Length (20-64)} (days) & \textbf{Avg. Lockdown (65+)} & \textbf{Length (65+)}  (days) & \textbf{Output Loss} (\%) & \textbf{Total Deaths} (\%) & \textbf{COVID-19 Deaths} (\%)\\
                \hline
        Benchmark& 0.3188 & 161 & 0.8819 & 207 & 7.3439 & 0.3266 & 0.2544 \\
         \hline
         \hline
         $\alpha_E = 0.21$ &0.3123 & 205 & 0.8878 & 239 & 8.9667 & 0.3265 & 0.2388\\
         $\alpha_E = 0.84$ & 0.2864 & 126 & 0.8663 & 167 & 5.2962 & 0.338 & 0.2848 \\
         \hline
         $\alpha_L = 0$ &0.3083 & 192 & 0.8863 & 227 & 8.3612 & 0.2456 & 0.2456\\
         $\alpha_L = 5 \times 10^{-5}$ & 0.2689 & 116 & 0.8511 & 154 & 4.6199 & 0.5319 & 0.2977 \\
         \hline
         $\alpha_I = 0$ & 0.3582 & 214 & 0.9057 & 249 & 10.6699 & 0.3712 & 0.2707 \\
         $\alpha_I = 6$ & 0.2173 & 144 & 0.7758 & 219 & 4.5236 & 0.2651 & 0.2106 \\
         $\alpha_I = 8$ &  0.1503 & 84 & 0.4962 & 188 & 1.9019 & 0.259 & 0.2328 \\
         $\alpha_I = 10$ & 0.041 & 24 & 0.3852 & 189 & 0.1558 & 0.245 & 0.2314 \\
         \hline
         $h = 0$ & 0.2874 & 126 & 0.8631 & 168 & 8.8592 & 0.338 & 0.2846 \\
         $h = 0.6$ & 0.3106 & 201 & 0.8866 & 235 & 5.842 & 0.3251 & 0.2394\\
         \hline
        $\rho = 0.5$ & 0.2989 & 193 & 0.8418 & 233 & 8.118 & 0.2753 & 0.1955 \\
        $\rho = 1$ & 0.3316 & 147 & 0.891 & 194 & 7.0292 & 0.3657 & 0.2975 \\
        \hline
        $\alpha_F = 0$ & 0.3065 & 175 & 0.855 & 211 & 7.6457 & 0.3324 & 0.2588 \\
        \hline
         $\nu = 1$ &0.3177 & 160 & 0.8803 & 206 & 6.8007 & 0.3264 & 0.2548 \\
         $\nu = 1.5$ &  0.3171 & 158 & 0.88 & 204 & 6.0805 & 0.3262 & 0.2556\\
         \hline
        $\eta = 0$ & 0.3188 & 161 & 0.8819 & 207 & 7.3439 & 0.3266 & 0.2544 \\
        $\eta = 10^6$ & 0.3188 & 161 & 0.8819 & 207 & 7.3439 & 0.3266 & 0.2544 \\
        \hline
       $\theta = 0.6$ &  0.3552 & 129 & 0.8792 & 161 & 6.7354 & 0.3589 & 0.2992 \\
        $\theta = 0.85$ &  0.2736 & 266 & 0.8815 & 304 & 9.7995 & 0.3075 & 0.2023 \\
        \hline    
        $\sigma(0.65) $ & 0.2982 & 185 & 0.8973 & 242 & 7.8084 & 0.2948 & 0.2135 \\
         $\sigma(0.7)$ & 0.2983 & 186 & 0.8919 & 279 & 7.8496 & 0.2724 & 0.1852 \\
          $\sigma(0.75) $ &  0.2793 & 202 & 0.8633 & Vaccine & 7.967 & 0.2732 & 0.1444 \\
        $\sigma(0.8)$ & 0.2775 & 204 & 0.9909 & Vaccine & 7.9881 & 0.2834 & 0.1416 \\
         $\sigma(0.9)$ & 0.2754 & 206 & 1.0 & Vaccine & 8.0011 & 0.2832 & 0.1405 \\
        $\sigma(1)$ & 0.2754 & 206 & 1.0 & Vaccine & 8.0011 & 0.2832 & 0.1405  \\
        \hline  
    \end{tabular}
    \caption{Parameter Robustness Results (Note: lockdown for 65+ ends at herd immunity) \\
    Benchmark Parameters: $\chi = 10/r$, $r = 0.001\%$,  $\alpha_E = 0.42$, $h$ = 0.4, $\alpha_L = 10^{-5}$, $\alpha_I = 1$, $\rho = 0.75$, F = 1, $\theta = 0.75$, $\nu  = 0.67$, $\eta = 10$,  Death rates from Table \ref{tab:low-death}, Herd Immunity Threshold = 60\% }
    \label{tab:robust-table}
    }
\end{table}


\clearpage

\newpage
\bibliographystyle{alpha} 
\bibliography{06-21-20} 

\end{document}